\documentstyle[twoside,psfig]{article}
\catcode`\@=11
\long\def\@makefntext#1{
\protect\noindent \hbox to 3.2pt {\hskip-.9pt  
$^{{\eightrm\@thefnmark}}$\hfil}#1\hfill}		

\def\thefootnote{\fnsymbol{footnote}}
\def\@makefnmark{\hbox to 0pt{$^{\@thefnmark}$\hss}}	
	
\def\ps@myheadings{\let\@mkboth\@gobbletwo
\def\@oddhead{\hbox{}
\rightmark\hfil\eightrm\thepage}   
\def\@oddfoot{}\def\@evenhead{\eightrm\thepage\hfil
\leftmark\hbox{}}\def\@evenfoot{}
\def\sectionmark##1{}\def\subsectionmark##1{}}

\oddsidemargin=\evensidemargin
\renewcommand{\thefootnote}{\fnsymbol{footnote}}

\newcounter{sectionc}\newcounter{subsectionc}\newcounter{subsubsectionc}
\renewcommand{\section}[1] {\vspace{12pt}\addtocounter{sectionc}{1} 
\setcounter{subsectionc}{0}\setcounter{subsubsectionc}{0}\noindent 
	{\tenbf\thesectionc. #1}\par\vspace{5pt}}
\renewcommand{\subsection}[1] {\vspace{12pt}\addtocounter{subsectionc}{1} 
	\setcounter{subsubsectionc}{0}\noindent 
	{\bf\thesectionc.\thesubsectionc. {\kern1pt \bfit #1}}\par\vspace{5pt}}
\renewcommand{\subsubsection}[1] {\vspace{12pt}\addtocounter{subsubsectionc}{1}
	\noindent{\tenrm\thesectionc.\thesubsectionc.\thesubsubsectionc.
	{\kern1pt \tenit #1}}\par\vspace{5pt}}
\newcommand{\nonumsection}[1] {\vspace{12pt}\noindent{\tenbf #1}
	\par\vspace{5pt}}

\newcounter{appendixc}
\newcounter{subappendixc}[appendixc]
\newcounter{subsubappendixc}[subappendixc]
\renewcommand{\thesubappendixc}{\Alph{appendixc}.\arabic{subappendixc}}
\renewcommand{\thesubsubappendixc}
	{\Alph{appendixc}.\arabic{subappendixc}.\arabic{subsubappendixc}}

\renewcommand{\appendix}[1] {\vspace{12pt}
        \refstepcounter{appendixc}
        \setcounter{figure}{0}
        \setcounter{table}{0}
        \setcounter{lemma}{0}
        \setcounter{theorem}{0}
        \setcounter{corollary}{0}
        \setcounter{definition}{0}
        \setcounter{equation}{0}
        \renewcommand{\thefigure}{\Alph{appendixc}.\arabic{figure}}
        \renewcommand{\thetable}{\Alph{appendixc}.\arabic{table}}
        \renewcommand{\theappendixc}{\Alph{appendixc}}
        \renewcommand{\thelemma}{\Alph{appendixc}.\arabic{lemma}}
        \renewcommand{\thetheorem}{\Alph{appendixc}.\arabic{theorem}}
        \renewcommand{\thedefinition}{\Alph{appendixc}.\arabic{definition}}
        \renewcommand{\thecorollary}{\Alph{appendixc}.\arabic{corollary}}
        \noindent{\tenbf Appendix \theappendixc #1}\par\vspace{5pt}}
\newcommand{\subappendix}[1] {\vspace{12pt}
        \refstepcounter{subappendixc}
        \noindent{\bf Appendix \thesubappendixc. {\kern1pt \bfit #1}}
	\par\vspace{5pt}}
\newcommand{\subsubappendix}[1] {\vspace{12pt}
        \refstepcounter{subsubappendixc}
        \noindent{\rm Appendix \thesubsubappendixc. {\kern1pt \tenit #1}}
	\par\vspace{5pt}}

\newcommand{\textlineskip}{\baselineskip=13pt}
\newcommand{\smalllineskip}{\baselineskip=10pt}

\def\eightcirc{
\begin{picture}(0,0)
\put(4.4,1.8){\circle{6.5}}
\end{picture}}
\def\eightcopyright{\eightcirc\kern2.7pt\hbox{\eightrm c}} 

\newcommand{\copyrightheading}[1]
	{\vspace*{-2.5cm}\smalllineskip{\flushleft
	{\footnotesize International Journal of Modern Physics C #1}\\
	{\footnotesize $\eightcopyright$\, World Scientific Publishing
	 Company}\\
	 }}

\newcommand{\publisher}[2]{{\begin{center}\footnotesize\smalllineskip 
	Received #1\\
	Revised #2
	\end{center}
	}}

\def\abstracts#1#2#3{{
	\centering{\begin{minipage}{4.5in}\footnotesize\baselineskip=10pt
	\parindent=0pt #1\par 
	\parindent=15pt #2\par
	\parindent=15pt #3
	\end{minipage}}\par}} 

\def\keywords#1{{
	\centering{\begin{minipage}{4.5in}\footnotesize\baselineskip=10pt
	{\footnotesize\it Keywords}\/: #1
	\end{minipage}}\par}}

\newcommand{\bibit}{\nineit}
\newcommand{\bibbf}{\ninebf}
\renewenvironment{thebibliography}[1]
        {\frenchspacing
	 \ninerm\baselineskip=11pt
         \begin{list}{\arabic{enumi}.}
        {\usecounter{enumi}\setlength{\parsep}{0pt}     
	 \setlength{\leftmargin 12.7pt}{\rightmargin 0pt} 
         \setlength{\itemsep}{0pt} \settowidth
	{\labelwidth}{#1.}\sloppy}}{\end{list}}

\newcounter{itemlistc}
\newcounter{romanlistc}
\newcounter{alphlistc}
\newcounter{arabiclistc}

\newcommand{\fcaption}[1]{
        \refstepcounter{figure}
        \setbox\@tempboxa = \hbox{\footnotesize Fig.~\thefigure. #1}
        \ifdim \wd\@tempboxa > 5in
           {\begin{center}
        \parbox{5in}{\footnotesize\smalllineskip Fig.~\thefigure. #1}
            \end{center}}
        \else
             {\begin{center}
             {\footnotesize Fig.~\thefigure. #1}
              \end{center}}
        \fi}

\newcommand{\tcaption}[1]{
        \refstepcounter{table}
        \setbox\@tempboxa = \hbox{\footnotesize Table~\thetable. #1}
        \ifdim \wd\@tempboxa > 5in
           {\begin{center}
        \parbox{5in}{\footnotesize\smalllineskip Table~\thetable. #1}
            \end{center}}
        \else
             {\begin{center}
             {\footnotesize Table~\thetable. #1}
              \end{center}}
        \fi}

\def\@citex[#1]#2{\if@filesw\immediate\write\@auxout
	{\string\citation{#2}}\fi
\def\@citea{}\@cite{\@for\@citeb:=#2\do
	{\@citea\def\@citea{,}\@ifundefined
	{b@\@citeb}{{\bf ?}\@warning
	{Citation `\@citeb' on page \thepage \space undefined}}
	{\csname b@\@citeb\endcsname}}}{#1}}

\newif\if@cghi
\def\cite{\@cghitrue\@ifnextchar [{\@tempswatrue
	\@citex}{\@tempswafalse\@citex[]}}
\def\citelow{\@cghifalse\@ifnextchar [{\@tempswatrue
	\@citex}{\@tempswafalse\@citex[]}}
\def\@cite#1#2{{$\null^{#1}$\if@tempswa\typeout
	{IJCGA warning: optional citation argument 
	ignored: `#2'} \fi}}

\def\pmb#1{\setbox0=\hbox{#1}
	\kern-.025em\copy0\kern-\wd0
	\kern.05em\copy0\kern-\wd0
	\kern-.025em\raise.0433em\box0}

\def\fnt#1#2{\footnotetext{\kern-.3em
	{$^{\mbox{\scriptsize #1}}$}{#2}}}

\def\ps@myheadings{%
    \let\@oddfoot\@empty\let\@evenfoot\@empty
    \def\@evenhead{\slshape\leftmark\hfil}
    \def\@oddhead{\hfil{\slshape\rightmark}}
    \let\@mkboth\@gobbletwo
    \let\sectionmark\@gobble
    \let\subsectionmark\@gobble
    }
\font\tenrm=cmr10
\font\tenit=cmti10 
\font\tenbf=cmbx10
\font\bfit=cmbxti10 at 10pt
\font\ninerm=cmr9
\font\nineit=cmti9
\font\ninebf=cmbx9
\font\eightrm=cmr8

\def\qed{\hbox{${\vcenter{\vbox{		    
   \hrule height 0.4pt\hbox{\vrule width 0.4pt height 6pt
   \kern5pt\vrule width 0.4pt}\hrule height 0.4pt}}}$}}

\renewcommand{\thefootnote}{\fnsymbol{footnote}}    

\def\bsc{{\sc a\kern-6.4pt\sc a\kern-6.4pt\sc a}}  	
\def\bflatex{\bf L\kern-.30em\raise.3ex\hbox{\bsc}\kern-.14em 
T\kern-.1667em\lower.7ex\hbox{E}\kern-.125em X} 

\begin{document}
\setlength{\textheight}{7.7truein}  

\thispagestyle{empty}

\markboth{\protect{\footnotesize\it K.N. Premnath and J. Abraham
}}{\protect{\footnotesize\it Turbulence modeling of jets in
discrete lattice BGK Boltzmann method}}


\normalsize\textlineskip

\setcounter{page}{1}

\copyrightheading{}			

\vspace*{0.88truein}

\centerline{\bf DISCRETE LATTICE BGK BOLTZMANN EQUATION}
\vspace*{0.035truein}
\centerline{\bf COMPUTATIONS OF TRANSIENT INCOMPRESSIBLE TURBULENT JETS}
\vspace*{0.37truein}
\centerline{\footnotesize KANNAN N. PREMNATH}
\baselineskip=12pt
\centerline{\footnotesize\it Maurice J. Zucrow Labs., School of Mechanical Engineering}
\baselineskip=10pt
\centerline{\footnotesize\it Purdue University, West Lafayette, Indiana 47907,
USA}
\centerline{\footnotesize\it E-mail: nandha@ecn.purdue.edu}

\vspace*{15pt}          
\centerline{\footnotesize JOHN ABRAHAM}
\baselineskip=12pt
\centerline{\footnotesize\it Maurice J. Zucrow Labs., School of Mechanical Engineering}
\baselineskip=10pt
\centerline{\footnotesize\it Purdue University, West Lafayette, Indiana 47907,
USA}
\centerline{\footnotesize\it E-mail: jabraham@ecn.purdue.edu}
\vspace*{0.225truein}
\publisher{(received date)}{(revised date)}

\vspace*{0.25truein}
\abstracts{In this paper, computations of transient, incompressible, turbulent, plane jets
using the discrete lattice BGK Boltzmann equation are reported. $\acute{A}$ priori derivation of the
discrete lattice BGK Boltzmann equation with a spatially and temporally dependent relaxation time parameter,
which is used to represent the averaged flow field, from its corresponding continuous form is given.
The averaged behavior of the turbulence field is represented by the standard
{\it k-$\epsilon$} turbulence model and computed using a finite-volume scheme on non-uniform
grids. Computed results are compared with analytical solutions, experimental data and results
of other computational methods. Satisfactory agreement is shown.}{}{}

\vspace*{5pt}


\vspace*{1pt}\textlineskip	
\section{Introduction}		
\vspace*{-0.5pt}
\noindent

The computational method based on the lattice Boltzmann equation (LBE) is relatively new 
for fluid dynamics. It is part of the paradigm of simulating complex physical phenomena, in
particular fluid flows, that are based on the observation that the interactions of quasi-particles
represented by simple models could give rise to very complex emergent phenomena. In 1986, in the
seminal works of Frisch, Hasslacher and Pomeau$^1$ and  Wolfram$^2$, the lattice gas automaton (LGA)
was introduced to simulate the fluid behavior described by the Navier-Stokes equations. Their work
showed that the key to recovering hydrodynamics from such models is that the underlying lattice structure,
in which the particles are constrained to move and collide while obeying mass and momentum 
conservation laws, satisfy certain symmetry properties. The lattice Boltzmann equation (LBE) was
introduced by MacNamara and Zanetti$^3$ to overcome certain drawbacks of the LGA such as the presence
of statistical noise and the lack of Galilean invariance. A number of refinements were made, such as the 
simplification of the description of the collision of particle populations$^{4,5}$ by means of the well-known
Bhatnagar-Gross-Krook (BGK)  model$^6$ which resulted in a considerable simplification of the LBE.
In the above sense, in which the LBE represents the stream-and-collide picture of the particle populations,
it may be essentially considered as an extension of the LGA.

On the other hand, it was believed that the LBE could also be connected to the Boltzmann equation, a well
known kinetic equation in non-equilibrium statistical mechanics, that describes the evolution of the
particle populations in terms of the distribution functions. It was formally shown that the LBE can be
derived $\acute{a}$ priori from the Boltzmann equation when its continuous velocity space is considerably
simplified to a certain discrete velocity space$^{7-9}$. With the foundations of the kinetic theory,
many ideas pertinent to the Boltzmann equation have been extended to the LBE or the discrete lattice
BGK Boltzmann equation. The rapid development and the applications of the lattice Boltzmann method (LBM)
that involves the solution of the LBE has been documented in several review papers$^{10-14}$ and
monographs$^{15,16}$. In particular, the ability of the method to model physics at a smaller scale
makes it a potentially promising tool to simulate fluid flows involving interfaces such as multiphase
and multicomponent flows and other complex flows. Algorithmically, the method involves operations that
are explicit and local and hence naturally amenable for implementation on almost all types of parallel
computers. Moreover, in the case of incompressible flows, conventional computational methods typically
require the solution of a Poisson-type equation for the computation of pressure field$^{17,18}$ which
may be time consuming and only partly parallelizable$^{19}$; on the other hand, the pressure is always computed
locally through an equation of state in the LBM.

Ever since the beginning of the development of the LGA, the precursor to the LBE, there were
speculations$^{20}$ about its application to the simulation of turbulence, the most dominant form of
fluid flow that occurs in nature as well as engineering applications. Because of the inherent
limitations of the LGA, some of which were discussed earlier, it was noted that its computational
requirements would be far more demanding than that of conventional macroscopic methods.
However, with the introduction of the LBE, with its physical as well as computational advantages,
direct computation of turbulence based on the LBE became feasible. Martinez \emph{et al.}$^{21}$
computed decaying turbulence in a shear layer and assessed the accuracy of the LBE by comparison of
its results with results obtained by employing the spectral method. It was shown that the LBE is almost 
as fast as the spectral
method on a serial computer and that it may well be more efficient if parallel computing strategies
are utilized. Some notable work using the LBE aimed at understanding the physics of turbulence
are the studies on the enstrophy cascade range$^{22}$ and the energy inverse cascade range$^{11}$ in
two-dimensional forced turbulence, the study of generalized extended self-similarity in three-dimensional,
inhomogeneous shear turbulence$^{23}$ and the work on three-dimensional turbulent channel
flow$^{24}$. It is interesting to note that there has been a growing interest in the use of kinetic
theory based approaches involving certain other forms of the Boltzmann equation to represent
turbulence$^{25,26}$. While such studies, where the LBE or some other forms of the Boltzmann equation
were used to resolve all the length and time scales, are important from a fundamental point
of view, for high-Reynolds number flows of engineering interest some form of modeling within the
LBE framework is desirable.

Turbulence modeling efforts within the LBE framework are being pursued through two approaches:
$(i)$ modeling based on the strict kinetic-theoretic formulations; and, $(ii)$ modeling based on traditional
concepts for which extensive literature already exist. The first approach is a more recent one that
involves the application of the renormalization group (RG) analysis to the simplified form of the
Boltzmann equation, such as the LBE, to develop a model for turbulence$^{27}$. It was found
that this introduces \emph{low Knudsen regime closure}, a feature that is peculiar to the kinetic
equation, which the authors believe could potentially offer new physical insights as well
as alternative mathematical treatments when compared to the Navier-Stokes equations. As this
approach is still in its early stages, much work remains to be done.
While the issues related to the first approach continue to
be addressed, it is important to develop practical turbulence models for the LBE so that flows
of practical interest can be computed within the LBE framework. In this respect, traditional,
statistical averaged procedures to turbulence modeling have been extended to the LBE$^{28-30}$. 
Essentially, in this second approach, the relaxation time parameter that appear in the BGK model
for the collision term in the LBE is now considered to be a \emph{spatially and temporally dependent variable}, 
instead of a constant. As a result, the expression for viscosity that can be obtained from the 
Chapman-Enskog multiscale expansion$^{31}$ can be considered to be the sum of the laminar viscosity
of the fluid and a spatially and temporally dependent eddy viscosity, which can be modeled using any statistical
averaged approach. An interesting question is whether we can derive this discrete lattice BGK   
Boltzmann equation with spatially and temporally dependent relaxation time parameter $\acute{a}$ priori from its
corresponding continuous form. In the next section, we address this question. 

It is important to assess the accuracy of the LBE used in conjunction with a turbulence model for 
flows of practical interest. Recently, it has been used to compute turbulent flow over an airfoil$^{32}$.
In this paper, we consider another application, namely, the computation of turbulent jets.
Fluid jets are encountered in many engineering applications such as internal combustion engines,
gas turbine combustors, industrial spray systems and a variety of other situations. Such flows
are almost exclusively turbulent, with Reynolds number, $Re$, of the order of $10^5$ or greater.
Understanding the transient process of mixing is important in such situations. Direct simulations
of jets would be limited to relatively low Reynolds numbers, of the order of few thousands. Hence
it becomes imperative to use a model to represent turbulence and thereby compute its effect on the
resolved flow field. In this work, we use the standard {\it k-$\epsilon$} turbulence model$^{33}$
in conjunction with the LBE, such that the latter would represent the unsteady,
mean flow field behavior, to study transient incompressible plane jets. The time-marching nature of LBM
is naturally beneficial for this problem that is inherently transient in nature.

The rest of the paper is organized as follows: In the next section, i.e.~Section~$2$, an $\acute{a}$ priori derivation
of the discrete lattice BGK Boltzmann equation, with a spatially and temporally dependent relaxation time, from its
continuous version is provided. The analysis is an extension of that provided in works of He and Luo$^{7,8}$.
In Section~$3$, the LBE as applied to simulate incompressible flows is discussed. As it has been
shown that adequate resolution is important to study jets$^{34}$, it is
necessary to employ non-uniform lattice grids within the LBE framework. Hence the LBE
extension to non-uniform lattice grids is subsequently described in Section~$4$. This is followed by a
discussion in Section~$5$ on the representation of turbulence within the LBE framework. The hybrid
numerical scheme for the solution of the LBE and turbulence model and the computational conditions
for the turbulent jet are described in Sections~$6$ and $7$ respectively. In Section~$8$, the
computed results are compared with analytical solutions, measurements, and published results
from other computational methods. Finally, the paper closes with summary and conclusions in
Section~$9$.

\setcounter{footnote}{0}
\renewcommand{\thefootnote}{\alph{footnote}}

\section{Analysis}
\noindent
Consider the Boltzmann equation with the BGK form for the collision term, where the relaxation time
parameter is taken to be a spatially and temporally dependent variable
\begin{equation}
\frac{\partial f}{\partial t}+\mbox{\boldmath$\xi$}\cdot \mbox{\boldmath $\nabla$} f=
 -\frac{1}{\lambda(\mbox{\boldmath$x$},t)} \left( f-f^{eq}\right).
 \label{eq:be}
\end{equation}

Here $f({\mathbf{x}},\mbox{\boldmath$\xi$},t)$ is the single-particle distribution function, 
$\mbox{\boldmath$\xi$}$ is the
particle velocity, $\lambda({\mathbf{x}},t)$ is the spatially and temporally dependent relaxation time, and $f^{eq}$ is 
the local Maxwellian given by
\begin{equation}
f^{eq}=\frac{\rho}{(2 \pi RT)^{D/2}}
	\mathrm{exp}\left[-\frac{\left(\mbox{\boldmath$\xi$}-\mathbf{u} \right)^2}
	{2RT} \right],
\label{eq:feq}
\end{equation}
where $R$ is the ideal gas constant, $D$ is the dimension of the space, and $\rho$, $\mathbf{u}$ and
$T$ are the \emph{mean} fluid density, velocity and temperature respectively. Limiting 
ourselves to the case of isothermal flows, the fluid density and velocity are obtained as the kinetic
moments of the distribution function, i.e.
\begin{eqnarray}
\rho & = &\int f d\mbox{\boldmath$\xi$} = \int f^{eq}d\mbox{\boldmath$\xi$},
\label{eq:rho}\\
\rho{\mathbf{u}} &=& \int \mbox{\boldmath$\xi$} f d\mbox{\boldmath$\xi$}
			   = \int \mbox{\boldmath$\xi$} f^{eq}d\mbox{\boldmath$\xi$}.
\label{eq:rhou}
\end{eqnarray}
Equation (\ref{eq:be}) can be formally written in the form of an inhomogeneous ordinary differential
equation with variable coefficients
\begin{equation}
\frac{df}{dt}+\frac{1}{\lambda\left({\mathbf{x}},t\right)}f=
\frac{1}{\lambda\left({\mathbf{x}},t\right)}f^{eq},
\end{equation}
where $\frac{d}{dt}= \frac{\partial}{\partial t}+\mbox{\boldmath$\xi$}\cdot \mbox{\boldmath $\nabla$}$ is the streaming
operator or the time derivative operator along the characteristic 
direction $\mbox{\boldmath$\xi$}$. The above equation can be formally integrated over a time step of
$\delta_t$, i.e.
\begin{eqnarray}
& &f\left( {\mathbf{x}}+\mbox{\boldmath$\xi$}\delta_{t},\mbox{\boldmath$\xi$},t+\delta_{t} \right)=
e^{-\int_{0}^{\delta_{t}}\frac{d{t'}}{\lambda\left( {\mathbf{x}}+{\mathbf{\xi}}{t'},t+{t'} \right)}}
\nonumber \\ & & \times \int_{0}^{\delta_t} 
e^{\int_{0}^{\delta_{t}}\frac{ds}{\lambda\left( {\mathbf{x}}+{\mathbf{\xi}}s,t+s \right)}}
\frac{1}{\lambda\left({\mathbf{x}}+\mbox{\boldmath$\xi$}t',t+{t'}\right)}
f^{eq}\left( {\mathbf{x}}+\mbox{\boldmath$\xi$}{t'},\mbox{\boldmath$\xi$},t+{t'}\right)dt'\nonumber\\&
+&
e^{-\int_{0}^{\delta_{t}}\frac{d{t'}}{\lambda\left( {\mathbf{x}}+{\mathbf{\xi}}{t'},t+{t'} \right)}}
f\left( {\mathbf{x}},\mbox{\boldmath$\xi$},t \right),
\label{eq:integ_f}
\end{eqnarray}
where $s,t' \in \left[0,\delta_t\right]$. Assuming that $\delta_t$ is small enough and 
the equilibrium distribution function, $f^{eq}$, and the relaxation time parameter, $\lambda$,
are locally smooth functions, the following linear approximation by Taylor expansion may be
made:
\begin{equation}
f^{eq}\left({\mathbf{x}}+\mbox{\boldmath$\xi$}t',\mbox{\boldmath$\xi$},t+t'\right)=
G_{1}+G_{21}t'+O\left( \delta_t^2 \right),
\end{equation}
\begin{equation}
\lambda \left({\mathbf{x}}+\mbox{\boldmath$\xi$}{t'},t+{t'} \right)=
\Lambda_{1}+\Lambda_{21}t'+O\left( \delta_t^2 \right),
\end{equation}
where
\begin{displaymath}
G_{1}=f^{eq}\left(\mathbf{x},\mbox{\boldmath$\xi$},t \right),
G_{21}=
\frac
{f^{eq}\left( {\mathbf{x}}+\mbox{\boldmath$\xi$}\delta_t,\mbox{\boldmath$\xi$},t+\delta_t\right)-
 f^{eq}\left({\mathbf{x}},\mbox{\boldmath$\xi$},t \right)}
 {\delta_t},
\end{displaymath}
\begin{displaymath}
\Lambda_{1}=\lambda\left( {\mathbf{x}},t \right),
\Lambda_{21}=
\frac
{\lambda\left( {\mathbf{x}}+\mbox{\boldmath$\xi$}\delta_t,t+\delta_t \right)-
 \lambda\left( {\mathbf{x}},t \right)}
 {\delta_t}.
\end{displaymath}
In the above, and in what follows, considering only first order accurate approximations in time$^{7,8}$, 
the leading terms of the order of $O(\delta_t^2)$ are neglected. Hence, the following approximation
in the integrand in Eq. (\ref{eq:integ_f}) may be made: 
\begin{displaymath}
\frac{1}{\lambda\left( {\mathbf{x}}+\mbox{\boldmath$\xi$}{t'},t+{t'} \right)}=
\frac{1}{\Lambda_{1}\left(1+\frac{\Lambda_{21}}{\Lambda_{1}}t' \right)}=
\frac{1}{\Lambda_{1}}\left(1-\frac{\Lambda_{21}}{\Lambda_{1}}t' \right)+
O\left( \delta_t^2 \right),
\end{displaymath}
\begin{displaymath}
\frac{1}{\lambda\left( {\mathbf{x}}+\mbox{\boldmath$\xi$}{t'},t+{t'} \right)}
f^{eq}\left( {\mathbf{x}}+\mbox{\boldmath$\xi$}t',\mbox{\boldmath$\xi$},t+t'\right)=
\frac{G_{1}+G_{21}t'}{\Lambda_{1}+\Lambda_{21}t'}=
\frac{1}{\Lambda_{1}}G_{1}+\
\Lambda_{1}\left( G_{1}-\frac{\Lambda_{21}}{\Lambda_{1}} \right)t'+
O\left( \delta_t^2 \right),
\end{displaymath}
\begin{displaymath}
e^{\int_{0}^{t'}\frac{ds}{\lambda\left( {\mathbf{x}}+{\mathbf{\xi}}s, t+s \right)}}=
e^{\int_{0}^{t'}\frac{1}{\Lambda_{1}}
\left( 1-\frac{\Lambda_{21}}{\Lambda_{1}}s \right)ds}
\approx
e^{\frac{t'}{\Lambda_{1}}}.
\end{displaymath}
Substituting the above in Eq. (\ref{eq:integ_f}), we get
\begin{eqnarray}
f\left( {\mathbf{x}}+\mbox{\boldmath$\xi$}\delta_{t},\mbox{\boldmath$\xi$},t+\delta_{t} \right)&=&
e^{-\frac{\delta_t}{\Lambda_{1}}}
\int_{0}^{\delta_t}e^{\frac{t'}{\Lambda_{1}}}
\left[ \frac{1}{\Lambda_{1}}G_{1}+
\frac{1}{\Lambda_{1}}\left(G_{1}-\frac{\Lambda_{21}}{\Lambda_{1}} \right)t'
\right] dt'\nonumber\\&\qquad+&e^{-\frac{\delta_t}{\Lambda_{1}}}
f\left( {\mathbf{x}},\mbox{\boldmath$\xi$},t \right).
\label{eq:integ_f1}
\end{eqnarray}
Now, as 
\begin{displaymath}
\int_{0}^{\delta_t}e^{\frac{t'}{\Lambda_{1}}}dt'=
\Lambda_{1}\left( e^{\frac{\delta_t}{\Lambda_{1}}}-1 \right),
\int_{0}^{\delta_t} t' e^{\frac{t'}{\Lambda_{1}}}dt'=
\Lambda_{1}\left[
\delta_t e^{\frac{\delta_t}{\Lambda_{1}}}-
\Lambda_{1}\left( e^{\frac{\delta_t}{\Lambda_{1}}}-1 \right)
\right],
\end{displaymath}
we can rewrite Eq.~(\ref{eq:integ_f1}) as follows:
\begin{eqnarray}
& &f\left( {\mathbf{x}}+\mbox{\boldmath$\xi$}\delta_{t},\mbox{\boldmath$\xi$},t+\delta_{t} \right)
\nonumber\\&=&
e^{-\frac{\delta_t}{\Lambda_{1}}}
\left\{
\frac{1}{\Lambda_{1}}G_{1}
\left[
\Lambda_{1}\left( e^{\frac{\delta_t}{\Lambda_{1}}}-1 \right)
\right]+
\frac{1}{\Lambda_{1}}
\left( G_{1}-\frac{\Lambda_{21}}{\Lambda_{1}}  \right)
\left[
\Lambda_{1}\left\{
\delta_t e^{\frac{\delta_t}{\Lambda_{1}}}-
\Lambda_{1}\left( e^{\frac{\delta_t}{\Lambda_{1}}}-1   \right)
	\right\}
	\right]
	\right\}
	\nonumber\\&\qquad+&e^{-\frac{\delta_t}{\Lambda_{1}}}
	f\left( {\mathbf{x}},\mbox{\boldmath$\xi$},t \right).
\end{eqnarray}
Employing Taylor expansion we get
$e^{\pm\frac{\delta_t}{\Lambda_{1}}}=1\pm\frac{\delta_t}{\Lambda_{1}}+O\left( \delta_t^2 \right)$.
Substituting this in the above equation
\begin{eqnarray}
f\left( {\mathbf{x}}+\mbox{\boldmath$\xi$}\delta_{t},\mbox{\boldmath$\xi$},t+\delta_{t} \right)&=&
\left( 1-\frac{\delta_t}{\Lambda_{1}} \right)
\left[
G_{1}\left( \frac{\delta_t}{\Lambda_{1}} \right)+
\frac{1}{\Lambda_{1}}
\left( G_{1}-\frac{\Lambda_{21}}{\Lambda_{1}}  \right)
\left( \frac{\delta_{t}^{2}}{\Lambda_{1}}   \right)
\right]\nonumber\\
&\qquad+&\left( 1-\frac{\delta_t}{\Lambda_{1}}  \right)
f\left( {\mathbf{x}},\mbox{\boldmath$\xi$},t \right),
\end{eqnarray}
and neglecting terms of the order of $O(\delta_t^2)$
\begin{equation}
f\left( {\mathbf{x}}+\mbox{\boldmath$\xi$}\delta_{t},\mbox{\boldmath$\xi$},t+\delta_{t} \right)=
\frac{\delta_t}{\Lambda_{1}}G_{1}
+\left( 1-\frac{\delta_t}{\Lambda_{1}}  \right)f\left( {\mathbf{x}},\mbox{\boldmath$\xi$},t \right),
\end{equation}
or, finally we obtain the time-discrete version of the Boltzmann equation  
\begin{equation}
f\left( {\mathbf{x}}+\mbox{\boldmath$\xi$}\delta_{t},\mbox{\boldmath$\xi$},t+\delta_{t} \right)-
f\left( {\mathbf{x}},\mbox{\boldmath$\xi$},t\right)=
-\frac{1}{\tau\left({\mathbf{x}},t\right)}
\left[
f\left( {\mathbf{x}},\mbox{\boldmath$\xi$},t\right)-
f^{eq}\left( {\mathbf{x}},\mbox{\boldmath$\xi$},t\right)
\right],
\label{eq:f_integ2}
\end{equation}
where $\tau \left( {\mathbf{x}},t \right)= \lambda \left( {\mathbf{x}},t \right)/\delta_t$. As shown 
by He and Luo$^{7,8}$, the equilibrium distribution function may be represented by a 
truncated small velocity expansion and the phase space is discretized such that the
numerical quadrature that is used in the calculation of the kinetic moments for the 
conserved variables is exact. Thus, for example, in the case of the two-dimensional, nine-velocity 
(D2Q9) model$^{4}$, shown in Fig.~$1$, the numerical quadrature naturally corresponds to the 
third-order Gauss-Hermite quadrature.
Hence, Eqs. (\ref{eq:feq}-\ref{eq:rhou}) and (\ref{eq:f_integ2}) may be
written in discretized form as
\begin{equation}
f_{\alpha}\left( {\mathbf{x}}+{\mathbf{e_{\alpha}}}\delta_{t},t+\delta_{t} \right)-
f_{\alpha}\left( {\mathbf{x}},t\right)=
-\frac{1}{\tau\left({\mathbf{x}},t\right)}
\left[
f_{\alpha}\left( {\mathbf{x}},t\right)-
f_{\alpha}^{eq}\left( {\mathbf{x}},t\right)
\right],
\label{eq:f_final}
\end{equation}
\begin{eqnarray}
\rho & = &\sum_{\alpha} f_{\alpha}\left({\mathbf{x}},t\right)
       =  \sum_{\alpha} f_{\alpha}^{eq}\left({\mathbf{x}},t\right) \label{eq:rho_final}, \\
       \rho{\mathbf{u}} & = &\sum_{\alpha} f_{\alpha}\left({\mathbf{x}},t  
       \right){\mathbf{e}}_{\alpha}
       =  \sum_{\alpha} f_{\alpha}^{eq}\left({\mathbf{x}},t  \right)
			{\mathbf{e}}_{\alpha},
\label{eq:rhou_final}
\end{eqnarray}
\begin{equation}
f_{\alpha}^{eq}=w_{\alpha}\rho
\left[
1+\frac{3}{c^2}\left({\mathbf{e}}_{\alpha}\cdot{\mathbf{u}} \right)+
  \frac{9}{2c^4}\left({\mathbf{e}}_{\alpha}\cdot{\mathbf{u}} \right)^2-
  \frac{3}{2c^2}u^2
\right],
\label{eq:feq_final}
\end{equation}
where
\begin{displaymath}
w_{\alpha}=
\left\{ \begin{array} {ll}
4/9  & {\alpha = 0}\\
1/9  & {\alpha = 1,2,3,4}\\
1/36 & {\alpha = 5,6,7,8}
\end{array} \right.
\end{displaymath}
and $\mathbf{e}_{\alpha}$ is the discrete set of the velocity space of $\mbox{\boldmath$\xi$}$ shown in
Fig.~$1$, with the Cartesian component of speed given by $c=\delta_x/\delta_t$ where $\delta_x$ is the
lattice spacing and $\alpha$ is the velocity direction. Also, 
$f_{\alpha}\left( {\mathbf{x}},t\right)=
w_{\alpha}f_{\alpha}\left( {\mathbf{x}},{\mathbf{e}}_{\alpha},t\right)$,
$f_{\alpha}^{eq}\left( {\mathbf{x}},t\right)=
w_{\alpha}f_{\alpha}^{eq}\left( {\mathbf{x}},{\mathbf{e}}_{\alpha},t\right)
$. 
Thus, the discrete lattice BGK Boltzmann equation with spatially and temporally dependent relaxation time parameter,
similar to the standard discrete lattice BGK Boltzmann equation with constant relaxation time parameter,
follows $\acute{a}$ priori from its corresponding continuous version and is independent of the LGA. It can
be shown using the Chapman-Enskog multiscale expansion$^{12,31}$ that in the long-wavelength 
limit, the mean density and velocity obey the unsteady Reynolds averaged Navier-Stokes (RANS) equations with
the viscosity related to the lattice parameters, i.e.  
\begin{equation}
\nu({\mathbf{x}},t)=\nu_{\mathrm{lam}}+\nu_{\mathrm{eddy}}({\mathbf{x}},t)=
\frac{1}{3}c^2\left( {\tau({\mathbf{x}},t})-\frac{1}{2}  \right)\delta_t,
\label{eq:viscosity}
\end{equation}
where $\nu_{\mathrm{lam}}$ is the kinematic viscosity of the fluid and $\nu_{\mathrm{eddy}}({\mathbf{x}},t)$
is the eddy viscosity which models the effect of turbulence on the flow field; The calculation of 
eddy viscosity is discussed in  
Section~$5$. In addition, it can be shown that the pressure is related to density by means of an equation
of state
\begin{equation}
p \left({\mathbf{x}},t\right)  = c^2_s \rho \left( {\mathbf{x}},t \right), 
\label{eq:pressure}
\end{equation}
where $c_s$ is the speed of sound, which is equal to $c/\sqrt{3}$.

\section{The Incompressible Model}
\noindent
It may be noted that the LBE as discussed above always simulates the \emph{weakly compressible} 
RANS equations which is valid for small Mach numbers, $Ma$. This is because of the fact that it models
physics \emph{locally}. Although true incompressibility, which amounts to infinite speed of sound,
cannot be achieved in this model, it can be modified such that it minimizes the compressibility 
effects to approximately represent incompressible flows. In this work, we employ the ``incompressible''
LBE model$^{35}$. According to this model, to approximately represent incompressible flows with a constant
density $\rho_{0}$, terms of the order of $o(Ma^2)$ in the formulation of the LBE are systematically 
eliminated. This leads to  a re-definition of the equilibrium distribution function, 
Eq.~(\ref{eq:feq_final}) and a modified expression for the calculation of fluid velocity 
$\rho_{0}{\mathbf{u}}=\sum_{\alpha}f_{\alpha}{\mathbf{e}}_{\alpha}$. We use the modified
equilibrium distribution function as derived by He and Luo$^{35}$: 
\begin{equation}
f_{\alpha}^{eq}=w_{\alpha}
\left[
\rho+\rho_{0}\left\{\frac{3}{c^2}\left({\mathbf{e}}_{\alpha}\cdot{\mathbf{u}} \right)+
\frac{9}{2c^4}\left({\mathbf{e}}_{\alpha}\cdot{\mathbf{u}} \right)^2-
\frac{3}{2c^2}u^2\right\}
\right],
\label{eq:feq_modified}
\end{equation}
where the coefficient $w_{\alpha}$ is the same as that used in Eq.~(\ref{eq:feq_final}).
In the literature, the LBE that results from these modifications is referred to as the ``incompressible'' 
LBE model.

\section{Non-uniform Lattice Grids}
\noindent
The use of non-uniform lattice grids is desirable in many applications and is important in the simulation
of jets where sharp gradients necessitate the use of high resolution in the near-orifice region$^{34}$. 
The original LBM is restricted to uniform grids, in that the minimum streaming distance
of the particle populations in one time step is exactly equal to the minimum lattice spacing. In other
words, the discretization of the configuration and particle velocity spaces are coupled.
This lockstep advection of particle population is a feature inherited from the LGA and is not
necessary for the LBE. Since the LBE is actually a simplified form of the Boltzmann equation which
can be solved without the coupling of the physical and the particle velocity spaces, it can be solved on
any mesh. Thus, it was proposed by He \emph{et al.}$^{36}$ that the collisions still take place on the
lattice grids in the manner discussed in the previous section; after collisions, the particle populations
move according to their velocities $\mathbf{e}_{\alpha}$; although the advected distance of the particle
populations may not, in general, coincide with the mesh spacing, the distribution functions in these
locations can always be computed using \emph{interpolation}; after interpolation, the collision and 
the streaming steps are repeated. It has been shown that, if the interpolation method is at least
of second order, the Navier-Stokes equations can still be recovered from the LBE$^{37}$. 

In this paper, we employ a second order Lagrange interpolation scheme to implement this
interpolation-supplemented LBE. Figure~$2$ illustrates the use of this method on a stretched 
non-uniform lattice mesh. 
For example, for the particle velocity direction $\alpha=1$, at locations
$(i,j)$ which corresponds to the lattice site $A$, $(i-1,j)$ and $(i-2,j)$, the collision step is
first computed. Then, the particle populations from these locations stream in the positive $i$ direction
to a distance $|\mathbf{e}_{\alpha}|\delta_t$, which may not, in general, be equal to the local mesh
spacing. After streaming, the distribution functions from these three new locations are interpolated to
obtain the value of the distribution at the location $A$. For direction $\alpha=3$, the lattice site
locations are taken from the negative $i$ direction as indicated at site $B$ in the figure. A similar procedure 
is adopted for the other particle directions. With this arrangement, the incompressible LBE model is used
to simulate jets.

\section{Turbulence Modeling}
\noindent
The standard two-equation {\it k-$\epsilon$} turbulence model is employed to represent the effects of 
the length and time scales in the turbulent flow. It was suggested earlier by Succi \emph{et al.}$^{28}$
that these equations might be solved within the LBE framework by creating two additional populations,
with components in the same directions as the particle distribution for each of the two scalar fields.
An alternative approach is to solve the  {\it k-$\epsilon$} equations using entirely different 
computational grids with an appropriate numerical scheme$^{30}$. Here, we consider this latter approach to
model the unresolved length and time scales in the turbulent, plane jets. The equations for the 
turbulent kinetic energy, $k$ and the dissipation rate, $\epsilon$ are given by
\begin{equation}
\rho_{0}
\left(
\frac{\partial k}{\partial t}+{\mathbf{u}}\cdot\mbox{\boldmath$\nabla$} k
\right) =
\frac{\partial}{\partial x_j}
\left[
\left(
\mu_{\mathrm{lam}}+\frac{\mu_T}{\sigma_k}
\right)
\frac{\partial k}{\partial x_j}
\right]+
\tau_{ij}S_{ij}-
\rho_{0}\epsilon,
\label{eq:k}
\end{equation}
\begin{equation}
\rho_{0}
\left(
\frac{\partial \epsilon}{\partial t}+{\mathbf{u}}\cdot\mbox{\boldmath$\nabla$} \epsilon
\right) =
\frac{\partial}{\partial x_j}
\left[
\left(
\mu_{\mathrm{lam}}+\frac{\mu_T}{\sigma_{\epsilon}}
\right)
\frac{\partial \epsilon}{\partial x_j}
\right]+
C_{\epsilon_1}\frac{\epsilon}{k}\tau_{ij}S_{ij}-
C_{\epsilon_2}\rho_{0}\frac{\epsilon^2}{k}.
\label{eq:eps}
\end{equation}
The Reynolds stress and strain rate tensors, $\tau_{ij}$ and $S_{ij}$ respectively, are related by
the linear constitutive relation through the Bousinessq approximation
\begin{equation}
\tau_{ij}=2 \mu_T S_{ij}-2/3\rho_{0}k\delta_{ij},
\label{eq:tau}
\end{equation}
and the eddy viscosity is given by
\begin{equation}
\nu_{\mathrm{eddy}}\left( {\mathbf{x}},t \right)=
\nu_T=
\frac{\mu_T}{\rho_{0}}=
C_{\mu}\frac{k^2}{\epsilon}.
\label{eq:eddy_viscosity}
\end{equation}
The values of the model coefficients, $C_{\epsilon_1}$,$C_{\epsilon_2}$,$\sigma_k$,$\sigma_{\epsilon}$,
$C_{\mu}$ used in this work are the same as that provided in Launder and Spalding$^{38}$. The strain 
rate tensor is directly
computed from the second kinetic moment of the non-equilibrium part of the distribution function, without
taking recourse to the finite-differencing of the velocity field, using   
\begin{equation}
S_{ij}=-\frac{3}{2}
	\frac{1}{\delta_t}
	\frac{1}{\tau\left({\mathbf{x}},t\right)}
	\sum_{\alpha}\frac{\mathrm{e}_{\alpha i} \mathrm{e}_{\alpha j}}{c^2}
	\frac{\left(f_{\alpha}-f_{\alpha}^{eq}  \right)}{\rho_{0}}.
\label{eq:strain_rate}
\end{equation}

\section{The Hybrid Computational Scheme}
\noindent
The mean flow field is computed using the LBE, Eq.~(\ref{eq:f_final}), supplemented by an
interpolation step, and the turbulence using the {\it k-$\epsilon$} model transport equations.
In this paper, we use the conservative formulation of
the {\it k-$\epsilon$} equations and solve them using a non-uniform \emph{finite-volume} (FV) based 
scheme. The computational mesh for the {\it k-$\epsilon$} equations is the same as that employed
for the lattice grids, with each finite-volume cell centered at a lattice site, e.g. the location
$P$ shown in Fig.~$3$. 
Following the procedure in Magi$^{17,18}$, the discretized equations 
are written in \emph{implicit} form with respect to such cells, with convective fluxes represented by
an upwind scheme and the diffusive fluxes obtained using central differences across the faces,
represented by the symbols $N$, $W$, $S$ and $E$ in Fig.~$3$.
They are solved using the \emph{strongly implicit procedure} (SIP) due to Stone$^{39}$ to obtain rapid 
convergence of the solutions.

\section{Computational Conditions}
\noindent
Fluid is injected in a plane domain at a constant velocity, $U_{\mathrm{inj}}$ with $U_{\mathrm{inj}}/c=0.1$,
where $c$ is the particle speed, through a slot of width $d$ such that $d/\delta_x=4$, where 
$\delta_x$ is the minimum lattice spacing, in a relatively large closed chamber. This choice of parameters
would keep the flow in the incompressible range so that the application of the incompressible LBE model 
becomes
valid. We consider the density of the injected fluid, $\rho_{0}$, to be equal to 1.0. The ambient fluid, which
is also of the same density, is assumed to be quiescent initially. The D2Q9 model of Fig.~$1$ is employed.
No slip boundary conditions are
imposed at the chamber walls. In this work, the boundary conditions are implemented using the 
non-equilibrium bounce back scheme suggested by Zou and He$^{40}$. For the {\it k-$\epsilon$}  
equations, the inlet conditions are specified, based on assuming equilibrium of turbulence production
and dissipation. Thus, the inlet turbulent kinetic energy and the dissipation rate are
$k_{\mathrm{in}}=1.5{u'}_{\mathrm{in}}^2$ and $\epsilon_{\mathrm{in}}=k_{\mathrm{in}}^{1.5}/l_d$
respectively. Here, ${u'}_{\mathrm{in}}$ is the root mean square value of the fluctuating component
of velocity and $l_d$ is the integral length scale. We assume ${u'}_{\mathrm{in}}=0.01U_{\mathrm{inj}}$
and $l_d = 0.25d$. Near the walls, wall functions
are used to specify the boundary conditions for the turbulent quantities$^{38}$. The Reynolds 
number of the jet, $Re_{\mathrm{d}}$, is based on the slot width and is defined by 
$Re_d = U_{\mathrm{inj}}d/\nu_{\mathrm{lam}}$. 

Two sets of problems are considered. First, to validate
the LBE without a turbulence model, computations of a laminar plane jet with a Reynolds number of $12$
are carried out and compared with similarity solutions. Second, we consider turbulent plane jet
with Reynolds number of $3\times10^4$ and compare the results with measurements and with those
obtained from other computational methods published in the literature. We are interested in determining 
the structure of the 
jet characterized by such quantities as the velocity distribution and the jet half-width, which will
be defined in the next section. The results will be reported in terms of lattice units, i.e. the
velocities are scaled by the particle velocity and the distance by the minimum lattice spacing.

\section{Results and Discussion}
\noindent
When the injected jet travels downstream of the slot, with $x$ as the axial distance from the
slot and $y$ as the transverse distance from the centerline of the jet, the centerline axial
velocity of the jet progressively decreases as a result of diffusion of momentum along the direction
transverse to the injected direction of the jet. Figure~$4$ shows the variation of the centerline
axial velocity, $U_{c}(x)$, of the laminar jet, normalized by the injection velocity, $U_{inj}$, 
as a function of the axial
distance from the slot, $x$, normalized by the slot width, $d$, at different times. It may be seen that
the velocity distribution is transient in character. Parts of the velocity profiles, nevertheless,
progressively approach steady state as time progresses. Thus, for instance, after $50,000$ time steps, 
the centerline velocity distribution up to at least $60$ slot widths downstream of the slot may be
considered to be steady. Now, the analytical solution for the structure of the jet$^{41,42}$ is 
applicable only for the steady part of the jet and hence they should only be used for comparison with the 
computed results
of the steady portion of the jet.
The analytical solution based on similarity considerations yields the axial velocity, $U(x,y)$, as a
function of the transverse distance. It is given by the expression 
$U(x,y)/U_{\mathrm{cl}} =\mathrm{sech}^2 \eta$, where $U_{\mathrm{cl}}(x)$ is the centerline
velocity and $\mathbf{\eta}$ is the similarity variable, a dimensionless coordinate given by
$\eta = 0.5\left( M_{\mathrm{inj}}/6\rho_{0}\nu_{\mathrm{lam}} \right)^{1/3} y/x^{2/3}$. Here,
$M_{\mathrm{inj}}$ is the injected momentum flow rate of the jet. The half-width of the jet, 
$y_{1/2}$, which is defined as the distance in the transverse direction from the centerline
where $U \left( x,y_{1/2} \right)=1/2U_{\mathrm{cl}}\left(x\right)$, also follows from the 
similarity solution as $y_{1/2}/d=A Re_{d}^{-2/3}\left(x/d\right)^{2/3}$, where $A$ is a constant
equal to $3.2038$. 

Figure~$5$ shows the variation of the computed half-widths at different times,
shown in symbols, as a function of the axial distance. 
Both axes are normalized by the slot width.
Also plotted on the same figure is the similarity solution. It may be seen that when
$t=40,000$, at least up to $15$ slot widths downstream of the slot the computed half-width profile
is steady. In this range, the growth of the half-width closely agrees with the analytical solution
within $4\%$.

The computed normalized velocity profile, $U_{\mathrm{cl}}(x,y)/U_{\mathrm{cl}}(x)$ as a function of the
similarity velocity variable, $\eta$ at $20$ slot widths downstream of the jet when $t=60,000$ is compared with
the similarity profile in Fig.~$6$. 
The computed and analytical results agree within $3\%$ for
$\eta \leq 1.5$, but the differences increase at distances greater than this, due to
wall effects.

Figure~$7$ shows the lattice/finite-volume mesh employed for the turbulent jet computations.  
A nominal resolution with $400$ lattice grids in both the axial and transverse directions are employed.
The grids are spaced non-uniformly with stretching applied in both the directions. The mesh
is considered to be uniform in the near-field region of the slot, where strong gradients in the 
flow field is expected. Effectively, the mesh accommodates a rectangular domain, whose length in
the axial distance equals $475d$ and that in the transverse direction equals $270d$. 

Figure~$8$ shows the velocity vector field in the domain of the turbulent jet after $10,000$  time steps. 
Also shown in the same figure is a plot representing the velocity vector field in the near-field of the
slot. It may be seen that a pair of vortices, the starting vortex or the head vortex pair, are 
formed from the two shear layers of the injected jet and the ambient fluid. As time progresses, it
was observed that these vortices convect downstream along with the unsteady part of the jet. Also
seen in the figure is the potential core, in which the jet preserves the injected velocity for
certain distance downstream from the slot. These basic feature of the jet are consistent with
those observed in experiments and in the numerical simulations based on the Navier-Stokes equations.

The computed axial velocity profile, normalized by the centerline velocity, is plotted as a function
of the transverse distance normalized by the jet half-width at $20$ slot widths downstream of the
slot for different times in Fig.~$9$. 
It is well known that the steady part of the transient jet
exhibits asymptotic self-similarity after a certain distance downstream of the slot. It may be
seen that at the downstream location plotted in Fig.~$9$, the velocity profiles at different
times almost overlap with one another, implying steady state there. Shown on the same plot in symbols
are the data from the measurements by Gutmark and Wygnanski$^{43}$. The computed results agree
within $10\%$ with the measured values.

As a  consequence of self-similarity, the axial variation of the centerline velocity and the
half-width of the jet obey the following scaling laws$^{42}$:
$U_{\mathrm{cl}} \left( x \right) \sim x^{-1/2}$ and $y_{1/2}\left( x \right) \sim x$.
Thus, in the region of self-similarity, $\left( U_{\mathrm{inj}}/U_{\mathrm{cl}}(x) \right)^2$
should be a linear function of the axial distance, $x$. Figure~$10$ shows the computed values of
$\left( U_{\mathrm{inj}}/U_{\mathrm{cl}}(x) \right)^2$, in filled circles, as a function of
the normalized axial distance after $80,000$ time steps. 
It may be seen that the computed results
indeed follow a linear variation as expected and agree qualitatively with the measurements.
Quantitatively, the data may be fitted to a linear curve that may be represented by
\begin{equation}
\left( \frac{U_{\mathrm{inj}}}{U_{\mathrm{cl}}(x)}\right )^2=
C_{1}\left[ \frac{x}{d}+C_{2} \right].
\label{Ccurvefit}
\end{equation}
In this work, this linear curve fit is applied to the computed results for the range of distances
given by $20 \leq x/d \leq 140$
and the constants $C_{1}$ and $C_{2}$ thus obtained are reported in Table~$1$, where data from 
other sources are  also compiled for comparison. 
The sources include the three measurements noted above
and the direct numerical simulation (DNS) data by Stanley \emph{et al.}$^{46}$ and the large eddy
simulation (LES) results by Le Ribault \emph{et al.}$^{47}$ It may be seen that there is a 
considerable scatter in the values of the constants. In particular, the constant $C_{2}$ which
represents the intercept of the linear fit vary significantly, which predominantly reflects the
quantitative difference between the computed variation and the measurement by 
Gutmark and Wygnanski$^{43}$. Physically, this constant is associated with the location of the
\emph{virtual origin} of the jet and hence it is expected that different methods could result
in different values of this constant.

Figure~$11$ shows the computed normalized half-width, in filled circles, plotted as a function
of the normalized axial distance. 
It may be seen that the computed results show a linear trend consistent with the similarity scaling law.
For a quantitative comparison, the results are fitted according to the
following linear curve
\begin{equation}
\left( \frac{y_{1/2}}{d}\right )=
K_{1}\left[ \frac{x}{d}+K_{2} \right].
\label{Kcurvefit}
\end{equation}
The constants $K_{1}$ and $K_{2}$ are presented in Table~$1$. The slope of this curve, i.e.
$dy_{1/2}/dx$ or $K_{1}$ is referred to as the \emph{spreading rate} of the jet and is a
constant in the similarity region. It is important to reproduce this quantity with sufficient
accuracy in engineering applications, as it is one of the measures of the rate of mixing of the 
injected and the ambient fluids. In Table~$1$, along with the computed spreading rate, the spreading
rate based on the \emph{steady} Reynolds averaged Navier-Stokes (RANS) equations with different two-equation
models as reported by Wilcox$^{33}$ and also by Magi \emph{et al.}$^{48}$ based on the \emph{unsteady}
RANS equations with the standard {\it k-$\epsilon$} are presented. The measured values for this quantity are 
in the range $0.100-0.110$. The values based on the direct numerical simulation (DNS) and that based 
on the turbulence models, including the large eddy simulation (LES) turbulence models, are in the 
range $0.090-0.110$. The computed value based on
the LBE/{\it k-$\epsilon$} FV scheme yields a value for the spreading rate as $0.0965$  which is
within $5\%$ agreement with these other studies.

\section{Summary and Conclusions}
In this paper, it is shown that the LBE with a spatially and temporally dependent relaxation time paper follows 
$\acute{a}$ priori from its continuous version. Computations of transient incompressible turbulent
plane jets are reported by employing this LBE in conjunction with the standard {\it k-$\epsilon$} 
turbulence model equations. The turbulence transport equations are solved using a finite-volume (FV)
scheme on non-uniform grids. The computed structure of the turbulent jet is
shown to be consistent with prior measurements and computed results. In particular, this hybrid
LBE-FV {\it k-$\epsilon$} scheme is found to reproduce the similarity scaling laws for turbulent jets,
i.e., the $x^{-1/2}$ decay for the centerline jet velocity and linear growth of the jet half-width. The
computed results agree with the measurements to within $10\%$. Laminar jet computations
are shown to be in agreement with the analytical solution. Although not shown here, we have shown 
elsewhere$^{49}$ that this hybrid approach is computationally more efficient on parallel computers as 
compared to the conventional schemes$^{19}$ when its inherent parallelism is exploited.

\nonumsection{References}

\eject

\begin{table}[htbp]
\tcaption{Comparison of the coefficients of the fitted linear curve for the normalized inverse
square axial velocity and the normalized axial variation of the half-width of the turbulent jet.}
\centerline{\footnotesize\smalllineskip
\begin{tabular}{l c c c c c}\\
\hline
Source     &   $Re_{d}$    &   $C_{1}$    &  $C_{2}$  &  $K_{1}$  & $K_{2}$\\
\hline
LBE / Std. $k-\epsilon$ model - this work        & 30,000 & 0.1371 &  0.0607 & 0.0965 & -0.488 \\
Unsteady RANS / Std. $k-\epsilon$ model$^{48}$   & ---    &   ---  &   ---   & 0.103  &  --- \\
Steady RANS / Std. $k-\epsilon$ model$^{33}$     & ---    &   ---  &   ---   & 0.108  &  --- \\
Steady RANS / Std. $k-\omega$ model$^{33}$       & ---    &   ---  &   ---   & 0.101  &  --- \\
LES/dynamic Smagorinsky SGS model$^{47}$         & 3,000  & 0.100  &  0.89   & 0.094  & 1.38 \\
LES/dynamic mixed SGS model$^{47}$               & 3,000  & 0.220  &  0.18   & 0.106  & 0.40 \\
DNS$^{46}$                                       & 3,000  & 0.201  &  1.23   & 0.092  & 2.63 \\
Measurement - Thomas \& Chu (1989)$^{45}$        & 8,300  & 0.220  &  -1.19  & 0.110  & 0.14 \\
Measurement - Browne \emph{et al.}(1983)$^{44}$  & 7,620  & 0.143  &  -9.00  & 0.104  & -5.00 \\
Measurement - Gutmark \& Wygnanski(1976)$^{43}$  & 30,000 & 0.123  &  4.47   & 0.100  & -2.00 \\
\hline\\
\end{tabular}}
\end{table}

\begin{figure}[htbp] 
\vspace*{13pt}
\centerline{\psfig{file=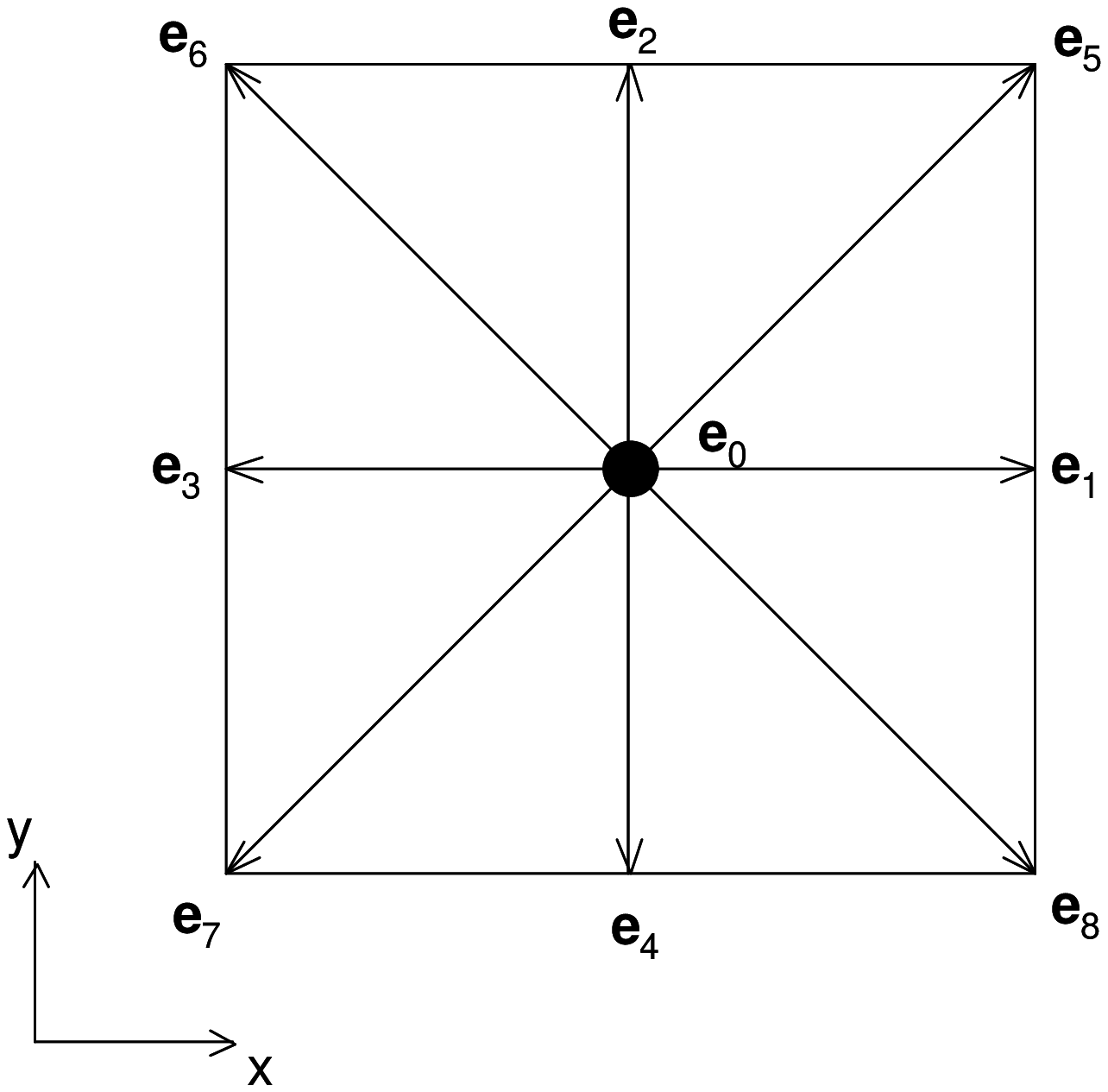}} 
\vspace*{13pt}
\fcaption{D2Q9 Lattice.}
\end{figure}

\begin{figure}[htbp] 
\vspace*{13pt}
\centerline{\psfig{file=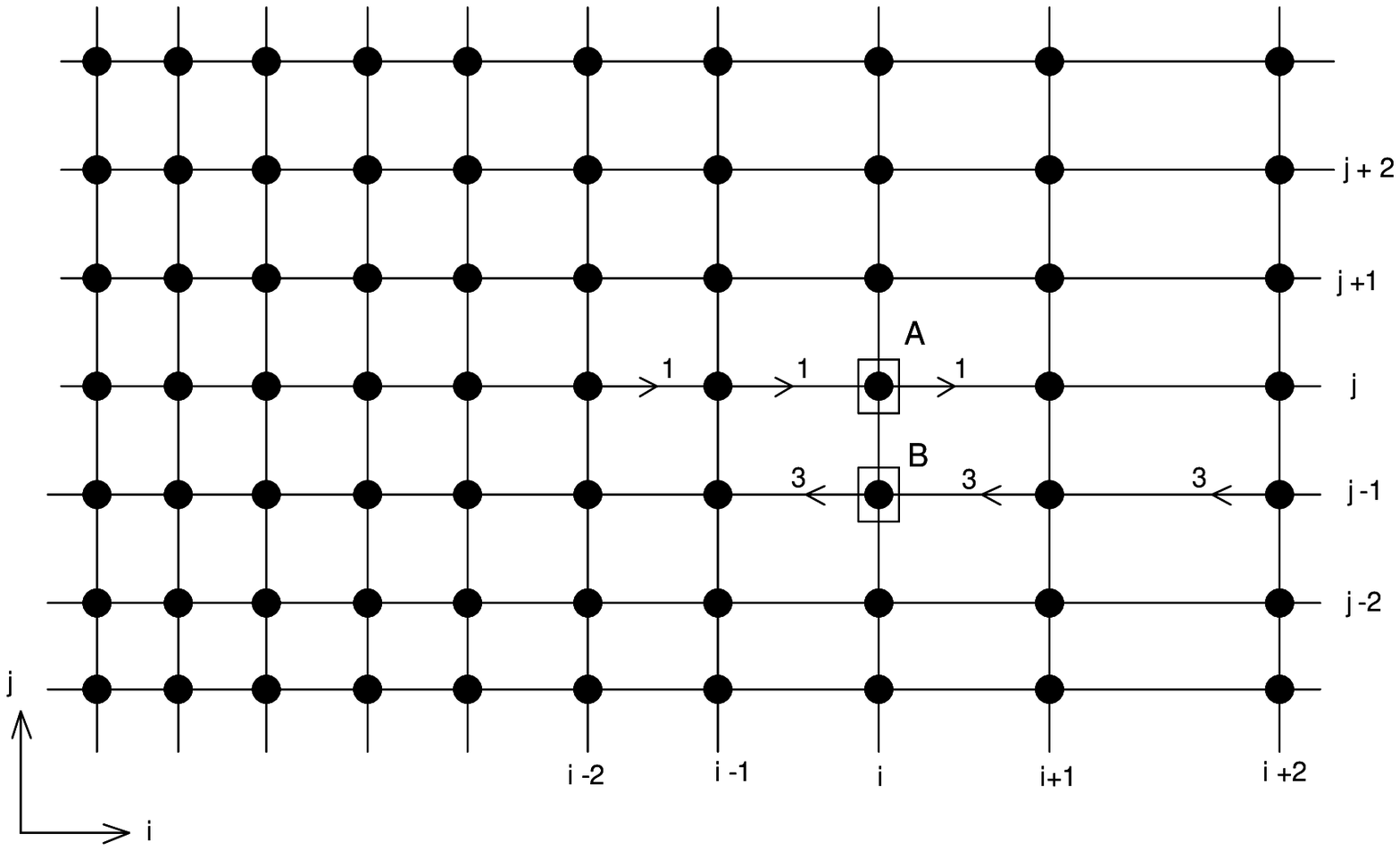}} 
\vspace*{13pt}
\fcaption{Non-uniform lattice mesh.}
\end{figure}

\begin{figure}[htbp] 
\vspace*{13pt}
\centerline{\psfig{file=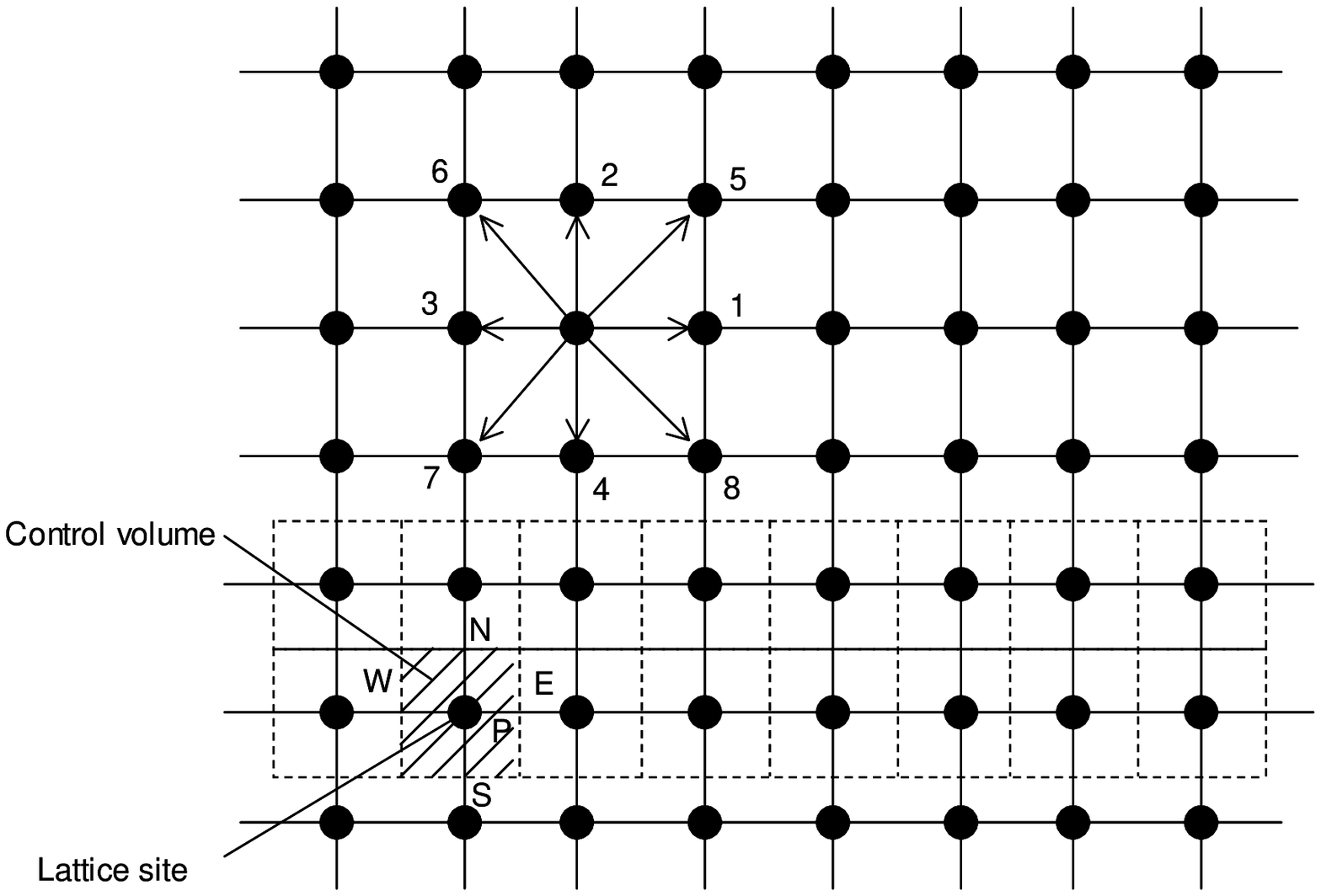}} 
\vspace*{13pt}
\fcaption{Finite control volumes embedded on lattice mesh.}
\end{figure}

\begin{figure}[htbp] 
\vspace*{13pt}
\centerline{\psfig{file=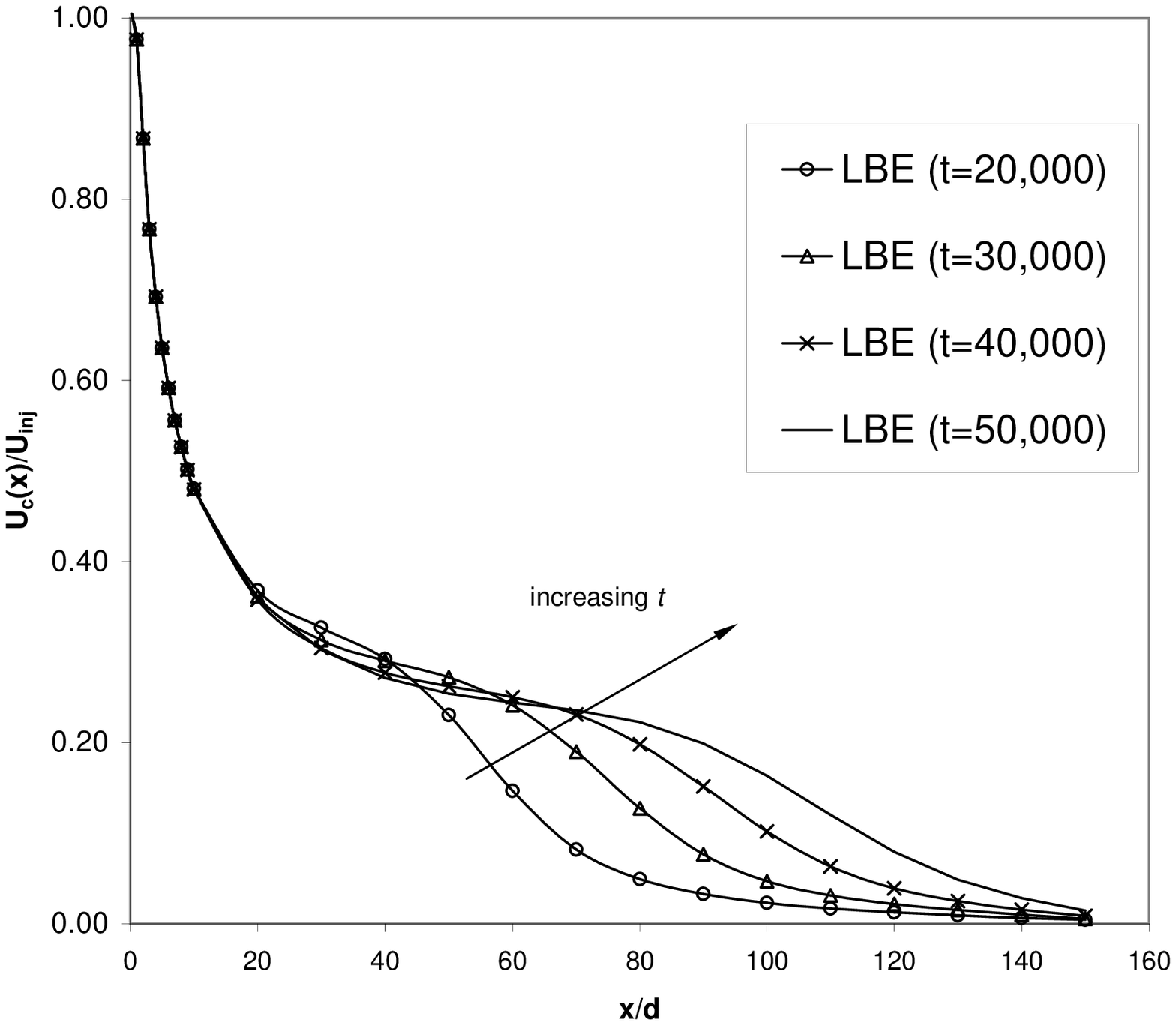}} 
\vspace*{13pt}
\fcaption{Transient axial velocity profiles as a function of axial distance of the laminar jet; $Re_{d}=12$.}
\end{figure}

\begin{figure}[htbp] 
\vspace*{13pt}
\centerline{\psfig{file=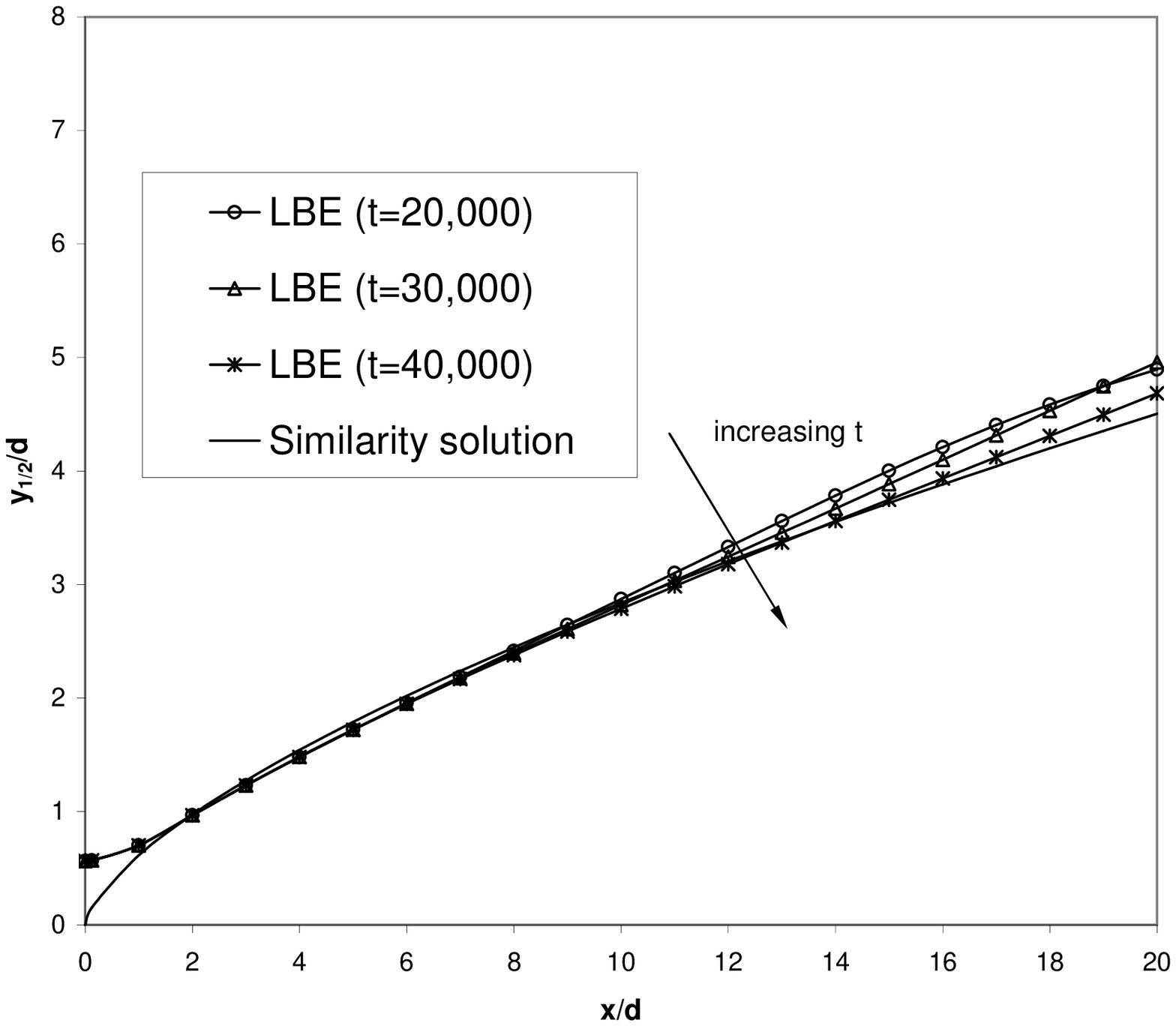}} 
\vspace*{13pt}
\fcaption{Comparison of computed results and similarity solution$^{41}$ for half-width of the laminar jet; $Re_{d}=12$.}
\end{figure}

\begin{figure}[htbp] 
\vspace*{13pt}
\centerline{\psfig{file=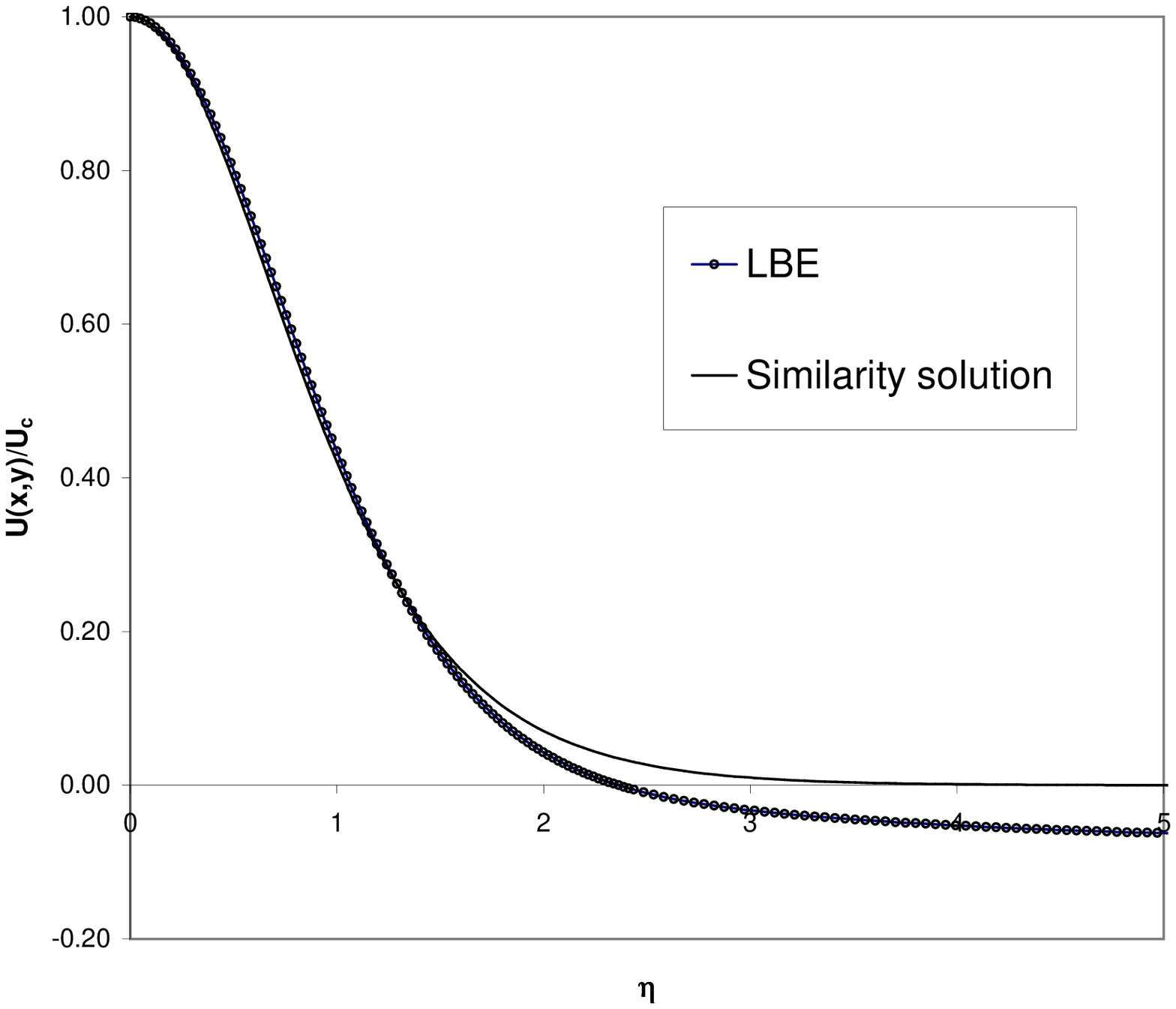}} 
\vspace*{13pt}
\fcaption{Comparison of computed results and similarity solution$^{41}$ for the normalized axial velocity
	  as a function of similarity variable of the laminar jet; $Re_{d}=12$, $x/d=20$.}
\end{figure}

\begin{figure}[htbp] 
\vspace*{13pt}
\centerline{\psfig{file=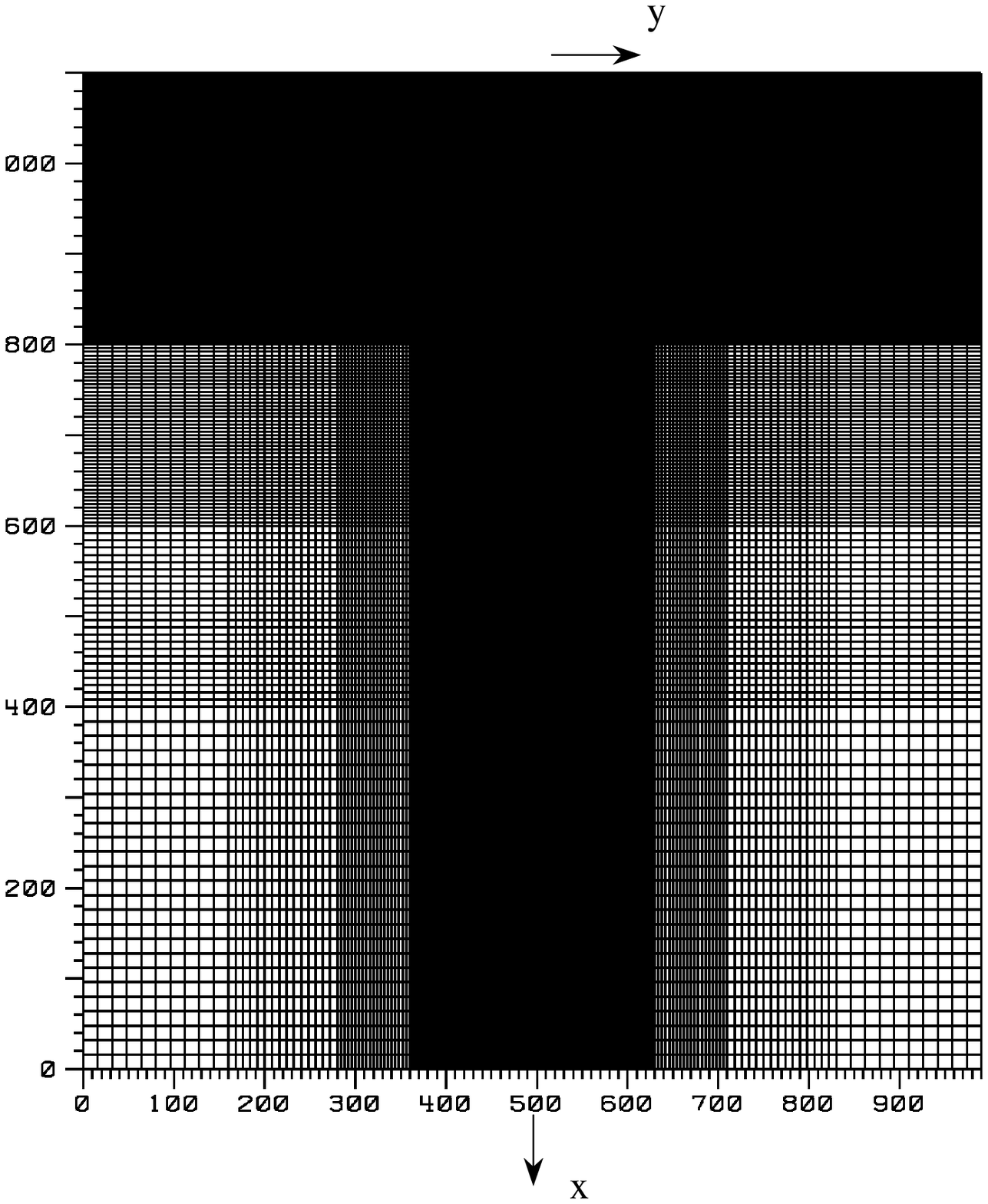}} 
\vspace*{13pt}
\fcaption{Computational lattice/finite volume mesh employed.}
\end{figure}

\begin{figure}[htbp] 
\vspace*{13pt}
\centerline{\psfig{file=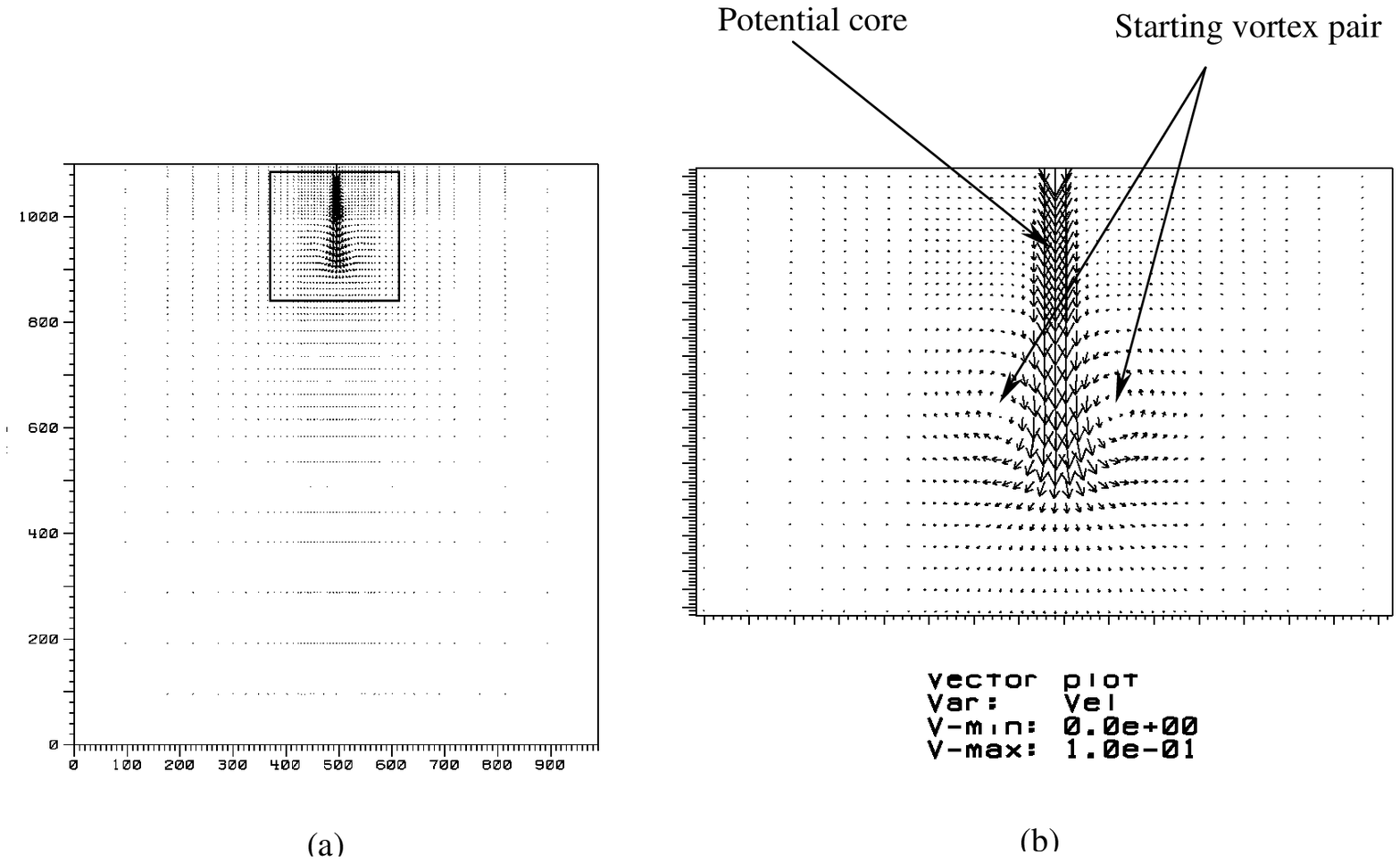}} 
\vspace*{13pt}
\fcaption{(a) Velocity vector field in the domain of the turbulent jet after $10,000$ time steps;
	  (b) Velocity vector field in the near-field of the slot showing the potential core and the
			starting vortex pair; $Re_{d}=30 \times 10^{4}$.}
\end{figure}

\begin{figure}[htbp] 
\vspace*{13pt}
\centerline{\psfig{file=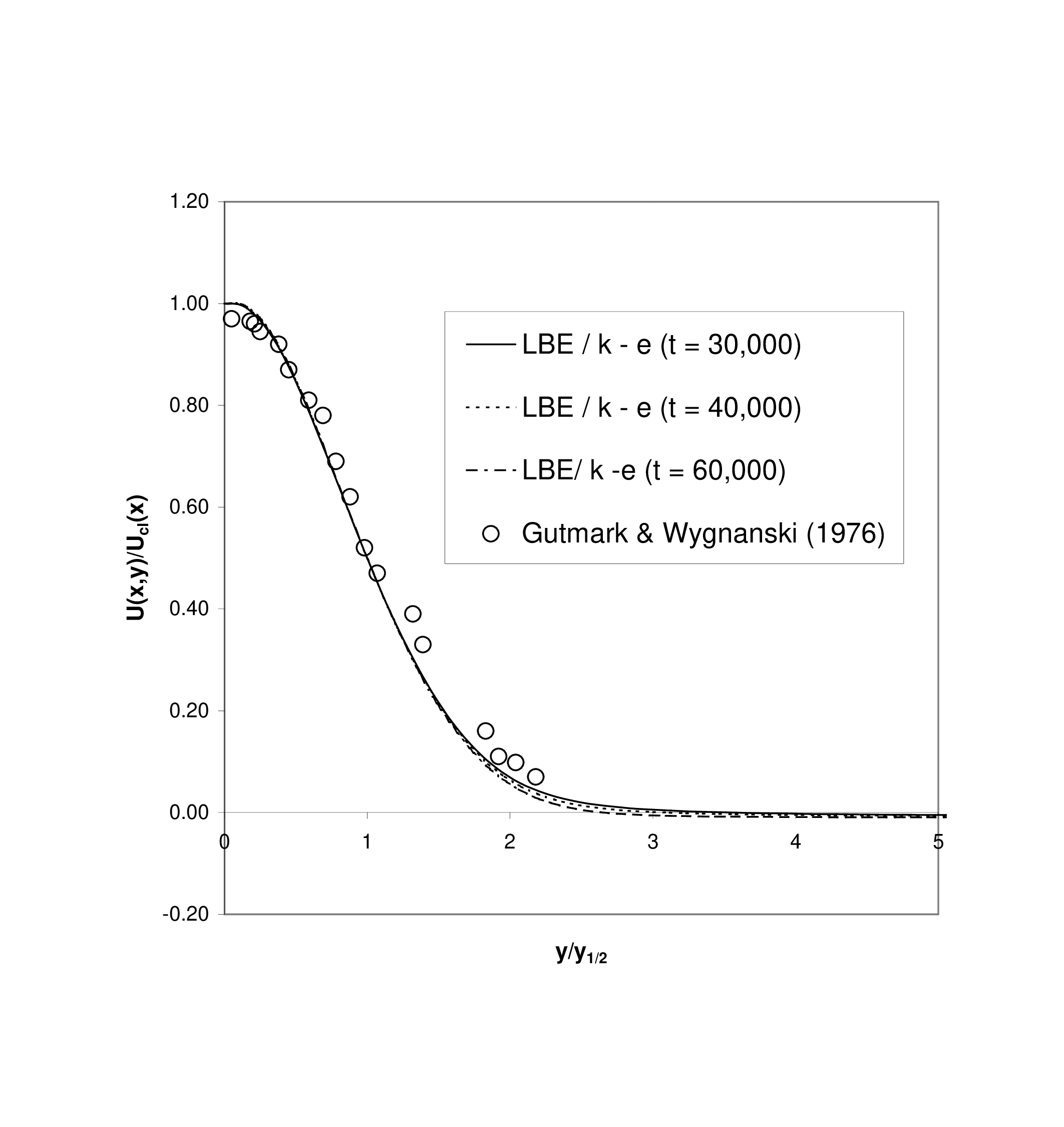}} 
\vspace*{13pt}
\fcaption{Comparison of computed results and measurements for the normalized axial velocity profile
	  as a function of the similarity variable of the turbulent jet;
          $Re_{d}=30 \times 10^{4}$, $x/d=20$.}
\end{figure}

\begin{figure}[htbp] 
\vspace*{13pt}
\centerline{\psfig{file=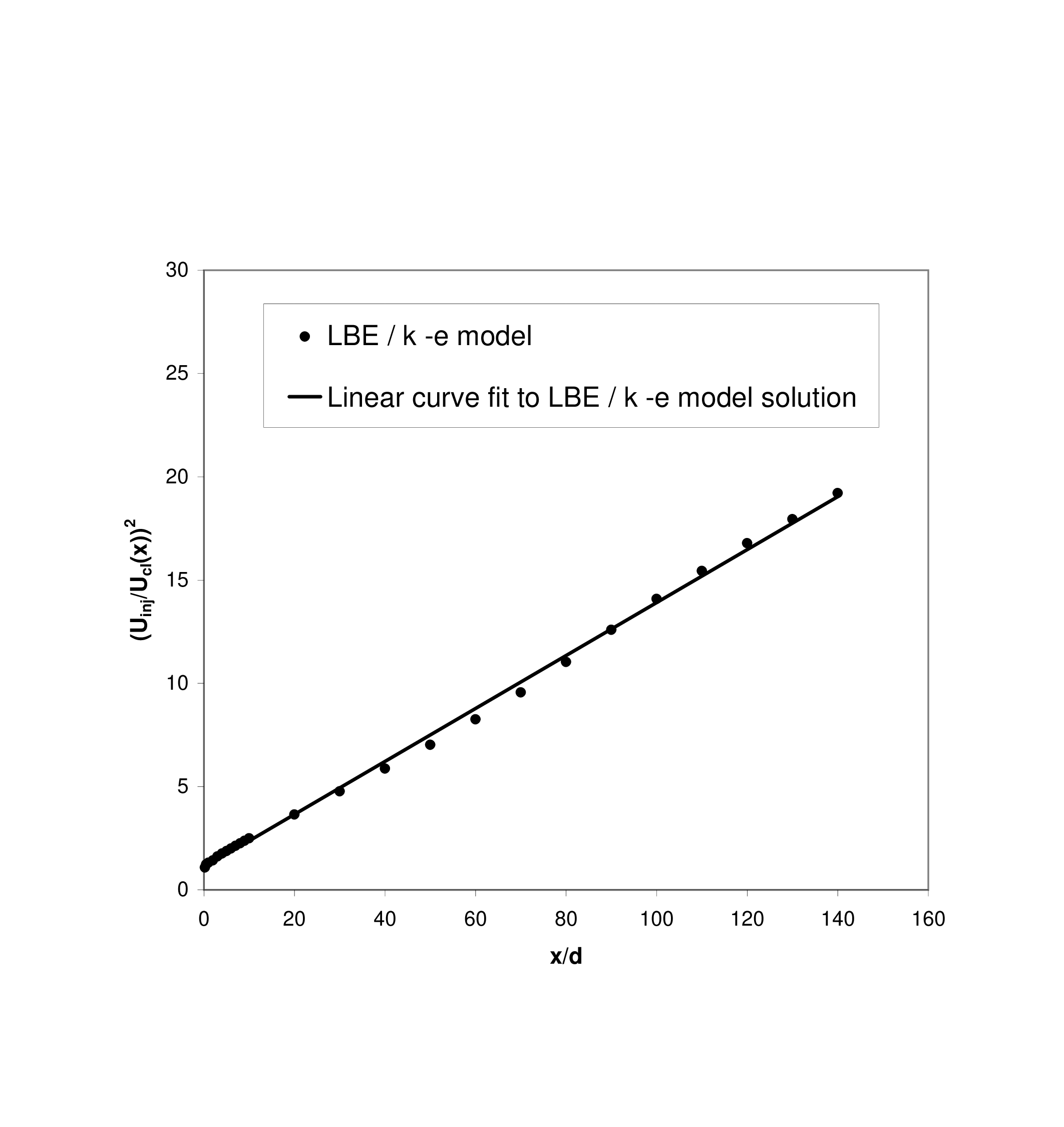}} 
\vspace*{13pt}
\fcaption{Normalized inverse square of the centerline velocity as a function of the normalized
	  axial distance of the turbulent jet; $Re_{d}=30 \times 10^{4}$.}
\end{figure}

\begin{figure}[htbp] 
\vspace*{13pt}
\centerline{\psfig{file=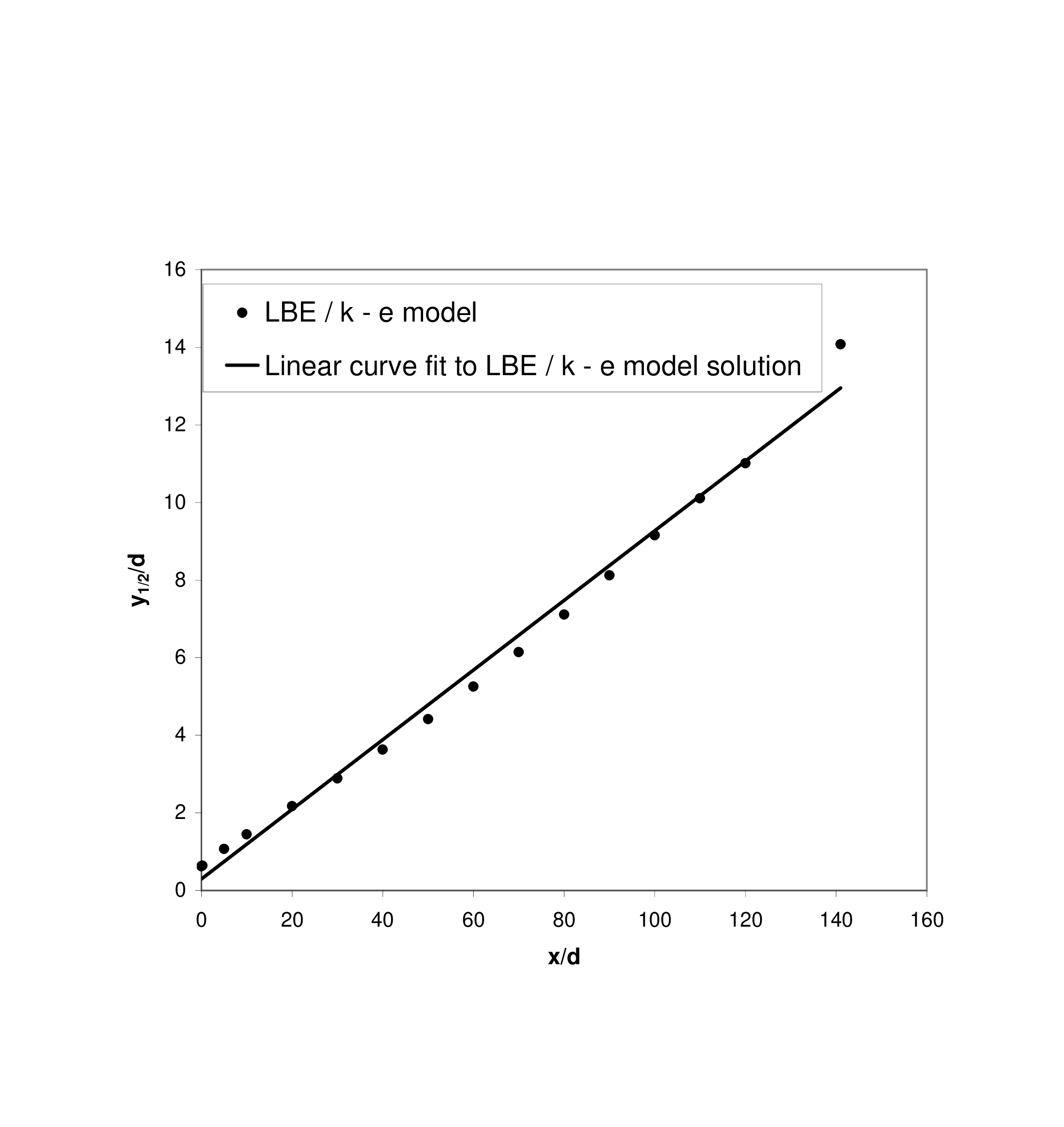}} 
\vspace*{13pt}
\fcaption{Normalized jet half-width as a function of the normalized axial distance
	  of the turbulent jet; $Re_{d}=30 \times 10^{4}$.}
\end{figure}


\noindent

\begin{thebibliography}{000}
\bibitem{1}
U. Frisch, B. Hasslacher, and Y. Pomeau {\bibit Phys. Rev. Lett.} {\bibbf 56}, 1505 (1986).

\bibitem{2}
S. Wolfram, {\bibit J. Stat. Phys.} {\bibbf 45}, 471 (1986).

\bibitem{3}
G.R. McNamara and G. Zanetti, {\bibit Phys. Rev. Lett.} {\bibbf 61}, 2332 (1988).

\bibitem{4}
Y. Qian, D.d'Humi$\grave{e}$res and P. Lallemand, {\bibit Europhys. Lett.} {\bibbf 17}, 479 (1992).

\bibitem{5}
H. Chen, S.Chen, D. Martinez and W. Matthaeus, {\bibit Phys. Rev. Lett.} {\bibbf 67}, 3776 (1991).

\bibitem{6}
P.L. Bhatnagar, E.P. Gross and M. Krook, {\bibit Phys. Rev.} {\bibbf 94}, 511 (1954).

\bibitem{7}
X. He and L.-S. Luo, {\bibit Phys. Rev.E} {\bibbf 55}, R6333 (1997).

\bibitem{8}
X. He and L.-S. Luo, {\bibit Phys. Rev.E} {\bibbf 56}, 6811 (1997).

\bibitem{9}
T. Abe, {\bibit J. Comp. Phys.} {\bibbf 131}, 241 (1997).

\bibitem{10}
R. Benzi, S. Succi and M.V. Vergasola, {\bibit Phys. Rep.} {\bibbf 222}, 147 (1992).

\bibitem{11}
Y.H. Qian, S. Succi and S.A. Orszag, {\bibit Annu. Rev. Comp. Phys.} {\bibbf 3}, 195 (1995).

\bibitem{12}
S. Chen and G.D. Doolen, {\bibit Annu. Rev. Fluid. Mech.} {\bibbf 30}, 329 (1998).

\bibitem{13}
S. Succi, O. Filippova, G. Smith and E. Kaxiras, {\bibit Comput. Sci. Eng.} {\bibbf 36}, 26 (2001).

\bibitem{14}
S. Succi, I.V. Karlin and H. Chen, {\bibit Rev. Mod. Phys.} {\bibbf 74}, 1203 (2002).

\bibitem{15} D. Wolf-Gladrow, {\it Lattice-Gas Cellular Automata and Lattice Boltzmann Models,
				   Lecture Notes in Mathematics, No. 1725}
				      (Springer, New York, 2000).

\bibitem{16} S. Succi, {\it The Lattice Boltzmann Equation for Fluid Dynamics and Beyond}
					 (Oxford University Press, 2001).

\bibitem{17}
F. Grasso and V. Magi,  in {\bibit Numerical Simulation of Combustion Phenomena,
                                   Lecture Notes in Physics} {\bibbf 241}, 282 (1985).

\bibitem{18} 
F. Grasso and V. Magi, in {\bibit Modern Research Topics in Aerospace Propulsion},
 ed. G. Angelino, L. De Luca and W.A. Sirignano (Springer-Verlag, 1991), p.~227.

\bibitem{19}
K. N. Premnath, V. Magi and J. Abraham, in {\bibit Grand Challenges in Computer Simulations ---
Proc. High Perf. Comput.  Symp.}, ed.  A. Tentner
(San Diego, CA, 2002), p.~3.

\bibitem{20}
S. A. Orszag and V. Yakhot, {\bibit Phys. Rev. Lett.} {\bibbf 56}, 1691 (1986).

\bibitem{21}
D.O. Martinez, W.Matthaeus, S. Chen and D. Montgomery, {\bibit Phys. Fluids} {\bibbf 6}, 1285 (1994).

\bibitem{22}
R. Benzi and S. Succi, {\bibit J. Phys. A: Math. Gen.} {\bibbf 23}, L1 (1990).

\bibitem{23}
R. Benzi, M.V. Struglia and R. Tripiccione, {\bibit Phys. Rev. E} {\bibbf 53}, R5565 (1996).

\bibitem{24}
G. Amati, S. Succi and R. Benzi, {\bibit Fluid Dyn. Res.} {\bibbf 19}, 289 (1997).

\bibitem{25}
V.V. Aristov, in {\bibit Proc. $21^{st}$ Int. Symp. Rarefied Gas Dynamics}
, ed.  R. Brun
(Marseuille, 1999), p.~187.

\bibitem{26}
A. Sakurai and F. Takayama, {\bibit Phys. Fluids} {\bibbf 15}, 1282 (2003).

\bibitem{27}
H. Chen, S. Succi and S. A. Orszag, {\bibit Phys. Rev. E} {\bibbf 59}, R2527 (1998).

\bibitem{28}
S. Succi, A. Amati and R. Benzi, {\bibit J. Stat. Phys.} {\bibbf 81}, 5 (1995).

\bibitem{29}
S. Hou, J. Sterling, S. Chen and G.D. Doolen, {\bibit Field Inst. Comm.} {\bibbf 6}, 151 (1996).

\bibitem{30}
C. Teixeira, {\bibit Int. J. Mod. Phys. C} {\bibbf 9}, 1159 (1998).

\bibitem{31} S. Chapman and T.G. Cowling, {\it Mathematical Theory of Non-Uniform Gases}
					 (Cambridge University Press, 1964).

\bibitem{32}
O. Filippova, S. Succi, F. Mazzocco, C. Arrighetti, G. Bella and D. Hanel, 
{\bibit J. Comp. Phys.} {\bibbf 170}, 812 (2001).

\bibitem{33} 
D. C. Wilcox, {\it Turbulence Modeling for CFD} (DCW Industries, CA,
1998).

\bibitem{34}
J. Abraham, {\bibit SAE Trans.} {\bibbf 106}, 147 (1997).

\bibitem{35}
X. He and L.-S. Luo, {\bibit J. Stat. Phys.} {\bibbf 88}, 927 (1997).

\bibitem{36}
X. He, L.-S. Luo and M. Dembo, {\bibit J. Comp. Phys.} {\bibbf 129}, 357 (1996).

\bibitem{37}
X. He, {\bibit Int. J. Mod. Phys. C} {\bibbf 8}, 737 (1997).

\bibitem{38}
B.E. Launder and D.B. Spalding, {\bibit Comput. Meth. Appl. Mech. and Engg.} {\bibbf 3}, 269 (1974).

\bibitem{39}
H. L. Stone, {\bibit SIAM J. Numer. Anal.} {\bibbf 5}, 530 (1968).

\bibitem{40}
Q. Zuo and X. He, {\bibit Phys. Fluids} {\bibbf 4}, 1510 (1997).

\bibitem{41}
W. G. Bickley, {\bibit Phil. Mag.} {\bibbf 23}, 727 (1937).

\bibitem{42} H. Schlichting, {\it Boundary Layer Theory}
					 (McGraw-Hill, New York, 1976).

\bibitem{43}
E. Gutmark and G. Wygnanski, {\bibit J. Fluid Mech.} {\bibbf 73}, 465 (1976).

\bibitem{44}
L.W.B. Browne, R.A. Antonia, S. Rajagopalan and A.J. Chambers, 
in {\bibit Structure of Complex Turbulent Shear Flow --- Proc. IUTAM Symp.}
, ed.  R. Dumas and L. Fulachier
(Marseuille, 1982), p. 411.

\bibitem{45}
F.O. Thomas and H.C. Chu, {\bibit Phys. Fluids A} {\bibbf 1}, 1556 (1989).

\bibitem{46}
S.A. Stanley, S. Sarkar and J.P. Mellado, {\bibit J. Fluid Mech.} {\bibbf 450}, 377 (2002).

\bibitem{47}
C. Le Ribault, S. Sarkar and S.A. Stanley, {\bibit Phys. Fluids} {\bibbf 11}, 3069 (1999).

\bibitem{48}
V. Magi, V. Iyer and J. Abraham, {\bibit Numer. Heat Transfer Part A} {\bibbf 40}, 317 (2001).

\bibitem{49}
K. N. Premnath and J. Abraham, in {\bibit Grand Challenges in Computer Simulations ---
Proc. High Perf. Comput.  Symp.}, ed.  A. Tentner
(San Diego, CA, 2002), p.~49.
\end{thebibliography}
\end{document}